\documentclass[%
 reprint,
 amsmath,amssymb,
 aps,pra,
]{revtex4-1}
\usepackage{adjustbox}
\usepackage{subfig}
\usepackage{color,hyperref}
\definecolor{darkblue}{rgb}{0.0,0.0,0.5}
\definecolor{darkred} {rgb}{0.5,0.0,0.0}
\hypersetup{
	linktocpage = true,						
	pdfpagelabels,
	pdfpagelayout = useoutlines,
	plainpages = false,
	hidelinks = false,
	bookmarks = false,
	bookmarksopen = false,
	bookmarksnumbered = false,
	breaklinks = true,
	unicode = false,
	pdftoolbar = true,
	pdfmenubar = true,
	pdffitwindow = true,
	pdfstartview = FitH, 
	pdfsubject = {Energy cascades for two interacting particles in a driven box},
	pdfauthor = {Sebastian Schiffer},
	pdfnewwindow = true,
	colorlinks = true,
	linkcolor = darkred,
	citecolor = darkblue,
	filecolor = magenta,
	urlcolor = darkblue,
	anchorcolor = green,
	hyperindex=true,
	hyperfigures
}	 
\usepackage{dsfont}
\usepackage{graphicx}
\usepackage{dcolumn}
\usepackage{bm}
\usepackage{xspace}
\usepackage[justification=RaggedRight, singlelinecheck=false]{caption}
\usepackage{tikz}
\usepackage{braket}
\usepackage{xcolor}

\usepackage{placeins}
\usepackage{float}

\newcommand{\ii}{\mathrm{i}}

\newcommand{\im}{\mathrm{Im}}
\newcommand{\re}{\mathrm{Re}}
\newcommand{\mc}{\mathcal}

\begin{document}

\title{Anderson localization transition in a  robust $\mathcal{PT}$-symmetric phase of a generalized Aubry-André model}

\author{Sebastian Schiffer}
 \email{sschiffer@swin.edu.au}
\author{Xia-Ji Liu}
 \author{Hui Hu}
 \author{Jia Wang}
  \email{jiawang@swin.edu.au}
\affiliation{
 Centre for Quantum Technology Theory, Swinburne University of Technology, Melbourne 3122, Australia
}

\date{\today}

\begin{abstract}
We study a generalized Aubry-Andr\'e model that obeys $\mathcal{PT}$-symmetry. We observe a robust $\mathcal{PT}$-symmetric phase with respect to system size and disorder strength, where all eigenvalues are real despite the Hamiltonian being non-hermitian. This robust $\mathcal{PT}$-symmetric phase can support an Anderson localization transition, giving a rich phase diagram as a result of the interplay between disorder and $\mathcal{PT}$-symmetry. Our model provides a perfect platform to study disorder-driven localization phenomena in a $\mathcal{PT}$-symmetric system.

\end{abstract}
\pacs{73.23.-b 72.70.+m 71.55.Jv 73.61.Jc 73.50.-h 73.50.T}
\maketitle

\emph{Introduction.}---Out-of-equilibrium open quantum systems are ubiquitous, where energy, particles, and information can transfer to or from the surrounding environment. In some limits, non-Hermitian Hamiltonians can well describe the quantum behavior of these systems \cite{SololovPLB1988, SololovNPA1989, RotterRPP1991, SololovAP1992, Carmichael1993, DittesPR2000, DaleyAP2014, YutoPRA2016, YutoNC2017, JinPRB2019}. The presence of complex eigenvalues of non-Hermitian Hamiltonians is a direct consequence of the non-preservation of probability due to loss and gain. However, non-Hermitian Hamiltonians that exhibit parity-time ($\mathcal{PT}$) symmetry can still possess a purely real spectrum, indicating the loss and gain are coherently balanced \cite{BenderPRL1998}. $\mathcal{PT}$-symmetry refers to the invariance of the Hamiltonian under a combined parity ($\mathcal{P}$) and time-reversal ($\mathcal{T}$) transformation, but not necessarily with $\mathcal{P}$ and $\mathcal{T}$ separately. Furthermore, a spontaneous $\mathcal{PT}$-symmetry breaking may occur when the degree of non-Hermiticity is large enough, where eigenvalues that come in complex conjugate pairs appear. We usually name the real (complex) spectral phase as a $\mathcal{PT}$-symmetric (-broken) phase.

$\mathcal{PT}$-symmetry became an active research area since the original work by Bender and Boettcher \cite{BenderPRL1998}. Applications of $\mathcal{PT}$-symmetry have been found in various physics areas, ranging from quantum field theories and mathematical physics \cite{BenderJMP1999, BenderPRL2002, BenderRPP2007, BenderPRL2017} to solid-state physics \cite{BendixPRL2009, JinPRA2009} and optics \cite{El-GanainyOL2007, MakrisPRL2008, MusslimaniPRL2008, LonghiPRL2009, LonghiPRB2009, GuoPRL2009}. It has recently attracted intense interest due to the rapid progress in atomic, molecular, and optical (AMO) experiments, where engineered loss and gain is accessible in a controllable manner \cite{GuoPRL2009, RuterNatPhys2010, RegensburgerNature2010, PengNatPhys2014, FengScience2014, HodaelScience2014, ZhenNature2015, ZeunerPRL2015, PoliNatComm2015, DopplerNature2016}. In particular, the real-to-complex spectral transition ($\mathcal{PT}$ transition) has been observed both in classical \cite{BenderAJP2013} and quantal systems \cite{LiNatComm2019}.

Another theoretical concept that has also gained a lot of attention recently thanks to experimental developments in photonic crystals \cite{NegroPRL2003, LahiniPRL2009, KrausPRL2012, VerbinPRL2013, VerbinPRB2015} and ultracold atoms \cite{RoatiNature2008, ModugnoRPP2010} is Anderson localization \cite{AndersonPR1958}. Anderson localization refers to the absence of a particle's diffusion induced by disorder. In a one-dimentional (1D) lattice model, an on-site cosine modulation incommensurate with the underlying lattice can be regarded as a highly correlated disorder, in a loose qualitative sense, and hence sometimes called quasi-disorder. Aubry and Andr\'e (AA) showed that a 1D tight-binding model with a quasi-disorder has a self-dual symmetry and manifests as a localization phase transition for all eigenstates at a critical modulation strength \cite{AubryAIPS1980}. This seminal work stimulated extensive theoretical and experimental investigations in various generalized AA models \cite{BiddlePRL2010, CaiPRL2013, DeGottardiPRL2013, GaneshanPRL2015, LiuPRB2015, WangPRB2016, CaoPRA2016, CestariPRB2016, ZengPRB2016, BaiPRA2018, Sanchez-PalenciaPRL2019,Sanchez-PalenciaPRL2020,GadwayPreprint2020}.

A localization transition can also occur in a non-Hermitian Hamiltonian system, such as non-Hermitian extensions of AA model \cite{ZengPRA2017, LiuPRB2020, zengPRB2020} and the Hatano-Nelson model with asymmetric hopping amplitudes \cite{HatanoPRL1996, HatanoPRB1997, HatanoPRB1998, HamazakiPRL2019}. A very recent study gives an interesting topological interpretation for the existence of the localization transition in the Hatano-Nelson model \cite{GongPRX2018}. However, whether an Anderson localization transition can exist in a $\mathcal{PT}$-symmetric Hamiltonian remains elusive. On the one hand, an exponential localization state induced by disorder requires a very large system size and can only be stable in the $\mathcal{PT}$-symmetric phase. On the other hand, an uncorrelated disorder usually does not respect $\mathcal{PT}$-symmetry, making the $\mathcal{PT}$-symmetric phase disappear for an arbitrarily weak disorder strength \cite{CristianPRA2015, JovicOL2012}. Even in a few studies that use an engineered $\mathcal{PT}$-symmetric disorder, the $\mathcal{PT}$-symmetric phase is still generally very fragile in the sense that it exists only for an exponentially small non-Hermicity parameter in the large system size limit \cite{BendixPRL2009, JoglekarPRA2010, LiangPRA2014, YucePLA2014}. Interestingly, the $\mathcal{PT}$-symmetric phase becomes robust if an asymmetric hopping is introduced, implying Anderson localization might exist \cite{JoglekarPRA2010, JoglekarPRA2011, ScottPRA2011, HarterPRA2016}.
 
\begin{figure}[b]
\includegraphics[scale=1]{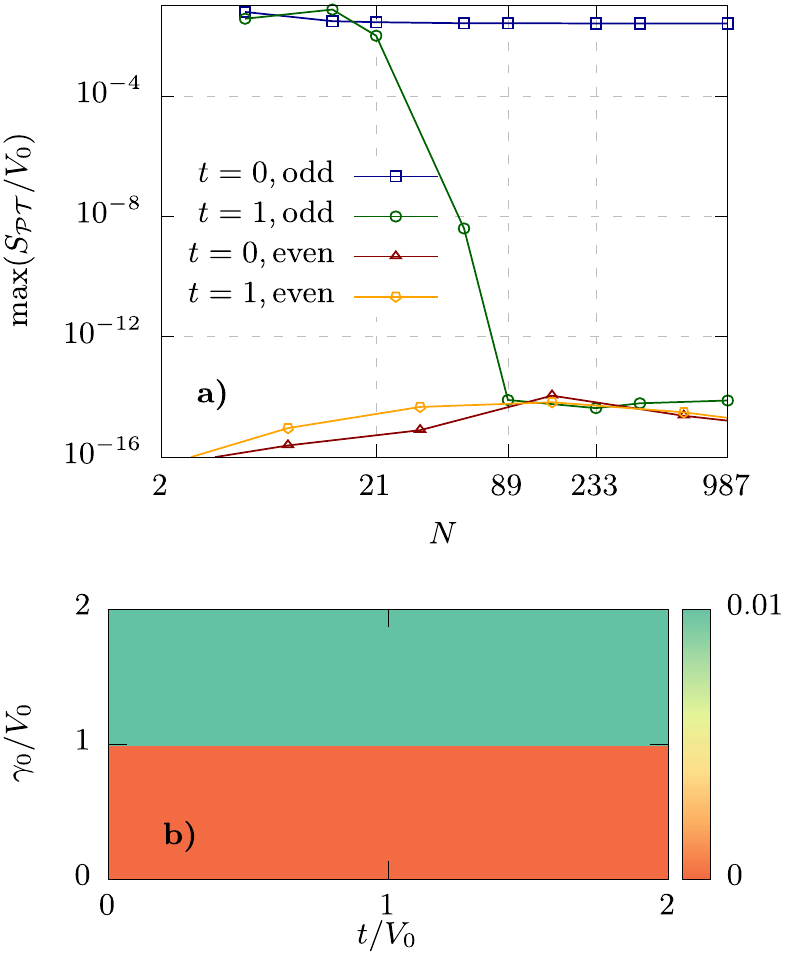}
\caption{(a) Maximum violation of the $\mathcal{PT}$-symmetry $\max(S_\mathcal{PT}/V_0)$ for even and odd chains at $\gamma_0=2$, $t=0$ and $t=1$. (b) $\langle I_\mathcal{PT} \rangle/V_0$ reveals the robust $\mathcal {PT}$-symmetric phase existing for $\gamma_0<1$ for arbitrary $t/V_0$.}\label{fig:Pic_PTbroken}
\end{figure}

\emph{Generalized AA Model.}---We study a generalized AA model with commensurate modulation in both on-site potentials and asymmetric imaginary hopping terms in this work. The Hamiltonian of the one-dimensional (1D) generalized AA model we consider here is given by
\begin{equation}
\hat H = \sum_{j=1}^N \left[ t_j \hat{c}_{j+1}^{\dagger}\hat{c}_{j}+ t_{j+1} \hat{c}_{j}^{\dagger}\hat{c}_{j+1} + V_j \hat{c}_{j}^{\dagger}\hat{c}_{j}\right],\label{eqn:H}
\end{equation}
where $\hat c_j^\dagger$ ($c_j$) is the creation (annihilation) operator at site $j$, and the subindex $j$ should be understood as $j \pmod{N}$. The on-site modulation is given by $V_j=2V_{0}\cos(2\pi\beta j+\varphi)$, and the hopping is complex and asymmetric: $t_j=t+\ii\gamma_{0}\sin(2\pi\beta j+\varphi)\ne t_{j+1}^*$. Here $V_0$ is the quasi-disorder strength, and $\gamma_{0}$ controls the non-Hermiticity. We also choose $\beta=M/N$, where $M$ and $N$ are two adjacent Fibonacci numbers, which are mutually prime. When $\gamma_{0}=0$, the model Hamiltonian reduces back to the traditional AA model with hopping amplitude $t$.
 
We analytically prove that this Hamiltonian is $\mathcal {PT}$-symmetric for a set of modulation phase factors $\varphi=\varphi_{\mathcal PT} \equiv m\pi/N$, where $m$ are odd (integer) numbers if $N$ is even (odd) \cite{Supp}. Surprisingly, we numerically observe that, under some conditions, the system's spectrum remains (up to the numerical accuracy) all real or complex-conjugate-paired for \emph{any arbitrary} $\varphi$. We test the violation of $\mathcal{PT}$-symmetry of our Hamiltonian by defining a measure that vanishes if all eigenenergies $E_k$ are either real or complex-conjugate-paired: 
\begin{align}
S_\mathcal{PT}&= \frac{1}{N}\sum_{k}^N|\im(E_k)|\prod_{m\neq k}^N\left[1-\delta(E_k,E_m)\right]\nonumber\\ 
&+\frac{1}{2N}\sum_{k}^N\sum_{m\neq k}^N\delta (E_k,E_m)|\im(E_k)+\im(E_m)|,
\end{align}
where $\delta(E_k,E_m)=1$ if the difference of the real parts is small enough i.e. $|\re(E_k)-\re(E_m)|<\epsilon_{\rm tol}$, and $0$ otherwise. We choose a tolerance $\epsilon_{\rm tol}=10^{-4}V_0$ for the numerical implementation. Here, we use $\re$ ($\im$) to denote the real (imaginary) part. Figure \ref{fig:Pic_PTbroken} (a) shows the behavior of the maximum of $S_\mathcal{PT}/V_0$ over $\varphi$ as a function of $N$ for some typical parameters to characterise whether the spectrum is purely real. Our numerical result shows that $S_\mathcal{PT}/V_0$ are always vanishingly small for even chains (i.e. $N$ is even). For long enough ($N>55$) odd chains, $S_\mathcal{PT}/V_0$ is also as small as the numerical precision except at the ray $\left\{t=0, \gamma_0>V_0\right\}$ in the $t$ - $\gamma_0$ parameter space, which is called ``special ray'' for convenience hereafter. We remark here that, for analyzing the disorder-driven localization transition, it is vital that the spectrum remains purely real or complex-conjugate-paired for arbitrary $\varphi$ since it allows us to average over the phase factor $\varphi$ to emulate disorder realization. Hereafter, unless specificed otherwise, we always average observables over $\varphi$ and denote the average as $\langle \cdot \rangle$, except at the special ray, where we only calculate for $\varphi=\varphi_{\mathcal PT}$.  

\emph{$\mathcal{PT}$-broken phase.}---For $\mathcal{PT}$-symmetric systems, the $\mathcal{PT}$-symmetry might be spontaneously broken, if the degree of non-Hermiticity is large enough \cite{BenderPRL1998}. In our system, we explore the parameter space to find both symmetry-broken and -unbroken regions. As the appearance of complex conjugate pairs in the spectrum of a $\mathcal{PT}$-symmetric system indicates the broken phase, we define a $\mathcal{PT}$-symmetry indicator as sum over the absolute values of the imaginary parts of the spectrum
\begin{align}
I_\mathcal{PT}=\sum_{k}\left|\im\left(E_{k}\right)\right|,\label{IPT}
\end{align}
which vanishes if the spectrum is purely real. We observe that $\langle I_\mathcal{PT} \rangle/V_0$ abruptly changes from finite to vanishingly small at the vicinity of $\gamma_{0} = V_{0}$ irrespective of the value of $t/V_0$, marking the boundary between ${\mathcal PT}$-symmetric and broken phase as depicted in Fig. \ref{fig:Pic_PTbroken}(b). The fact that a $\mathcal {PT}$-transition occurs at $\gamma_{0} = V_{0}$ for arbitrary $t/V_0$ implies the $\mathcal {PT}$-symmetric phase in our system is robust against strong disorder. We have also confirmed that this $\mathcal {PT}$-phase diagram is essentially unchanged for larger $N$, indicating the robustness against system size. The robustness of the $\mathcal {PT}$-symmetric phase in our system is in stark contrast to most of the previous studies, where the $\mathcal {PT}$-symmetric phase becomes exponentially fragile in the presence of disorder.
\begin{figure}[b]
	\includegraphics[scale=1]{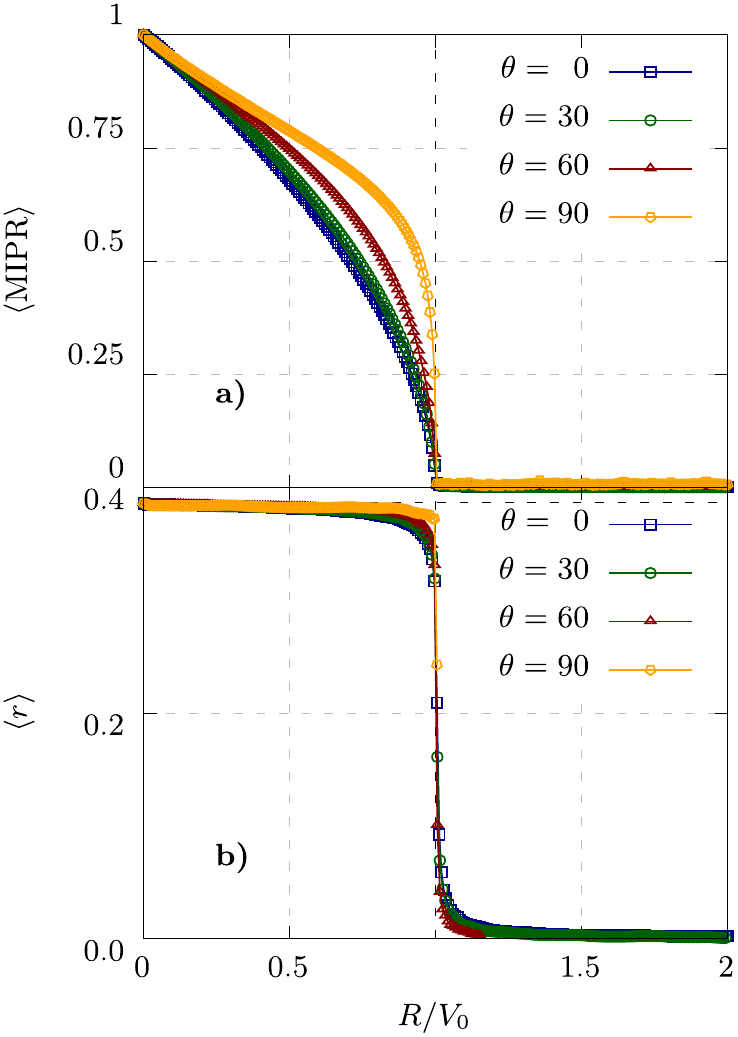}
	\caption{$\langle {\rm MIPR} \rangle$ and $\langle r \rangle$ as a function of $R=\sqrt{t^2+\gamma_0^2}$. a) shows the $\langle {\rm MIPR} \rangle$ at several different $\theta=\tan^{-1}(\gamma_0/t)$, indicating the localization transition occurs at $R=V_0$ for all $\theta$. b) shows the gap statistics $\langle r \rangle \approx 0.38$, the Poisson distribution value, in the strong disorder limit $t/V_0 \rightarrow 0$, and a rapid decay at the localization transition boundary.} \label{fig:Pic_Localization}
\end{figure}
 \begin{figure}[t]
 \includegraphics[scale=1]{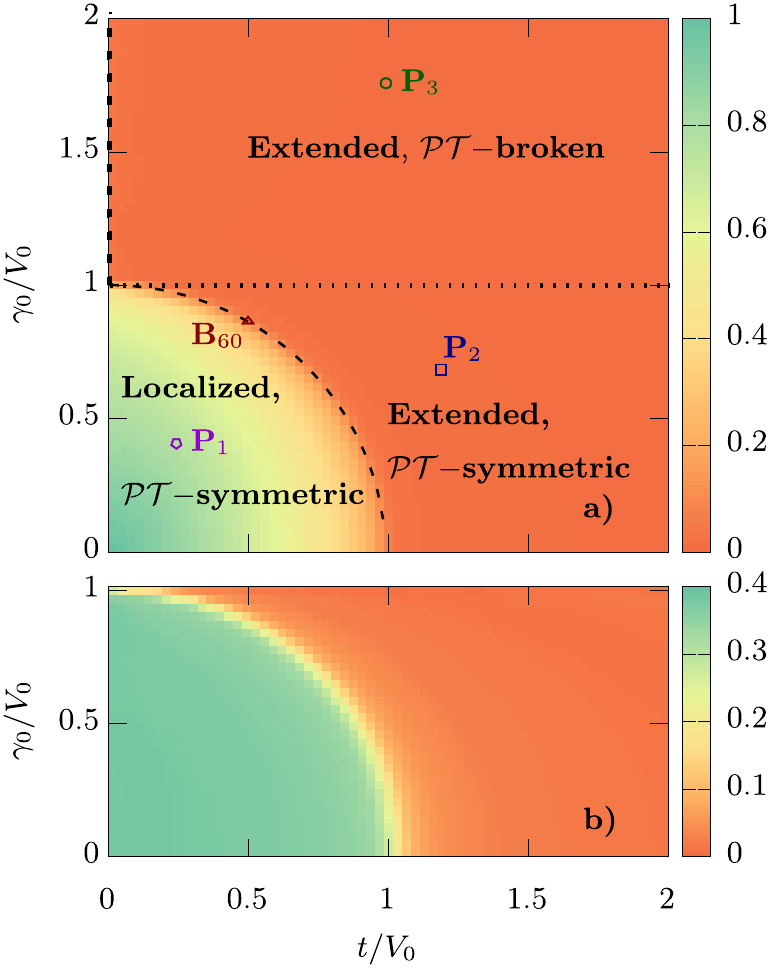}
    \caption{Phase diagrams of the system for $N=233$. a) and b) show $\langle {\rm MIPR} \rangle$ and gap statistics $\langle r \rangle$ respectively, both of which identify a localized phase within the quarter circle $\sqrt{t^2+\gamma_0^2}\leq V_0$. The localization-transition and $\mathcal {PT}$-transition boundaries are also indicated in a) by the thin dashed curve and the dotted line respectively. A thick dashed line illustrates the ``special ray'' $\{t=0, \gamma>1\}$ detailed in the main text. We also mark several specific points $\rm P_1$, $\rm P_2$, $\rm P_3$ and $\rm B_{60}$ in different phase regimes, which correspond to $\left\{t/V_0, \gamma_0/V_0 \right\} \approx \{0.24,0.42\}$, $\{1.2,0.69\}$, $\{1.0,1.74\}$ and $\{\cos(60^\circ),\sin(60^\circ)\}$. We exemplify properties of different phases on these points as detailed in the main text.}\label{fig:Pic_PhaseDiagram}
 \end{figure}

\emph{Localization.}---Next, we investigate the system for its localization behavior. A widely used measure for localization is the inverse participation ratio (IPR) \cite{Schaefer1980}. For a normalized wavefunction $\psi(j)$ of an hermitian Hamiltonian, the IPR is defined as the summation of the probability over all the sites $\sum_j p(j)^2 \equiv \sum_j |\psi(j)|^4$. In the case of non-hermitian Hamiltonians, the left and right eigenvectors can be orthonormalised in the sense that $\sum_j \psi_m^L(j)^* \psi_k^R(j) = \delta_{mk}$, where $p_k^{\rm LR}(j)=\psi_k^L(j)^* \psi_k^R(j)$ plays a similar role as probability at site $j$. Thus we define the IPR measure as \cite{Zhang2019}
\begin{align}
\operatorname{IPR}_{\rm LR}\left(E_{k}\right)=\left[\frac{\left(\sum_{j}\left|\psi_{k}^{L}(j)\psi_{k}^{R}(j)\right|\right)^{2}}{\sum_{j}\left|\psi_{k}^{L}(j)\psi_{k}^{R}(j)\right|^{2}}\right]^{-1},
\end{align}
which varies from being $\mathcal O (1/N)$ for eigenfunctions smeared uniformly over all sites to $\mathcal O (1)$ for those localized near a specific site. Therefore, the IPR can serve as an indicator for the localization transition. Averaging the IPR over all eigenfunctions and all quasi-disorder realizations gives the mean inverse participation ratio $\langle \textrm{MIPR}\rangle=\langle \sum_{k}\textrm{IPR}_{\rm LR}(E_k)/N \rangle$ \cite{Supp}. Figure \ref{fig:Pic_Localization} (a) shows the $\langle \textrm{MIPR}\rangle$ as a function of $R=\sqrt {t^2+V_0^2}$ for various $\theta=\textrm{atan}(\gamma_0/t)\in [0^\circ,90^\circ]$. These calculations are carried out for $N=1597$, where the numeric is well converged. The $\langle \textrm{MIPR} \rangle$ monotonically decreases from one to zero in the regime $R/V_0\in[0,1]$ and slower for larger $\theta$. The $\langle \textrm{MIPR}\rangle$ also essentially remains zero in the regime $R>V_0$ for any $\theta$. In the $t$ - $\gamma_0$ parameter space, $R/V_0$ can be recognised as the distance to the origin, and $\theta$ as the angle to the $t$-axis. Therefore, the localization boundary is located at the quarter circle arc $\sqrt{t^2+\gamma_0^2}=V_0$, which is also illustrated in the phase diagram in Fig. \ref{fig:Pic_PhaseDiagram} a).

\begin{figure*}[t]	
\includegraphics[scale=1]{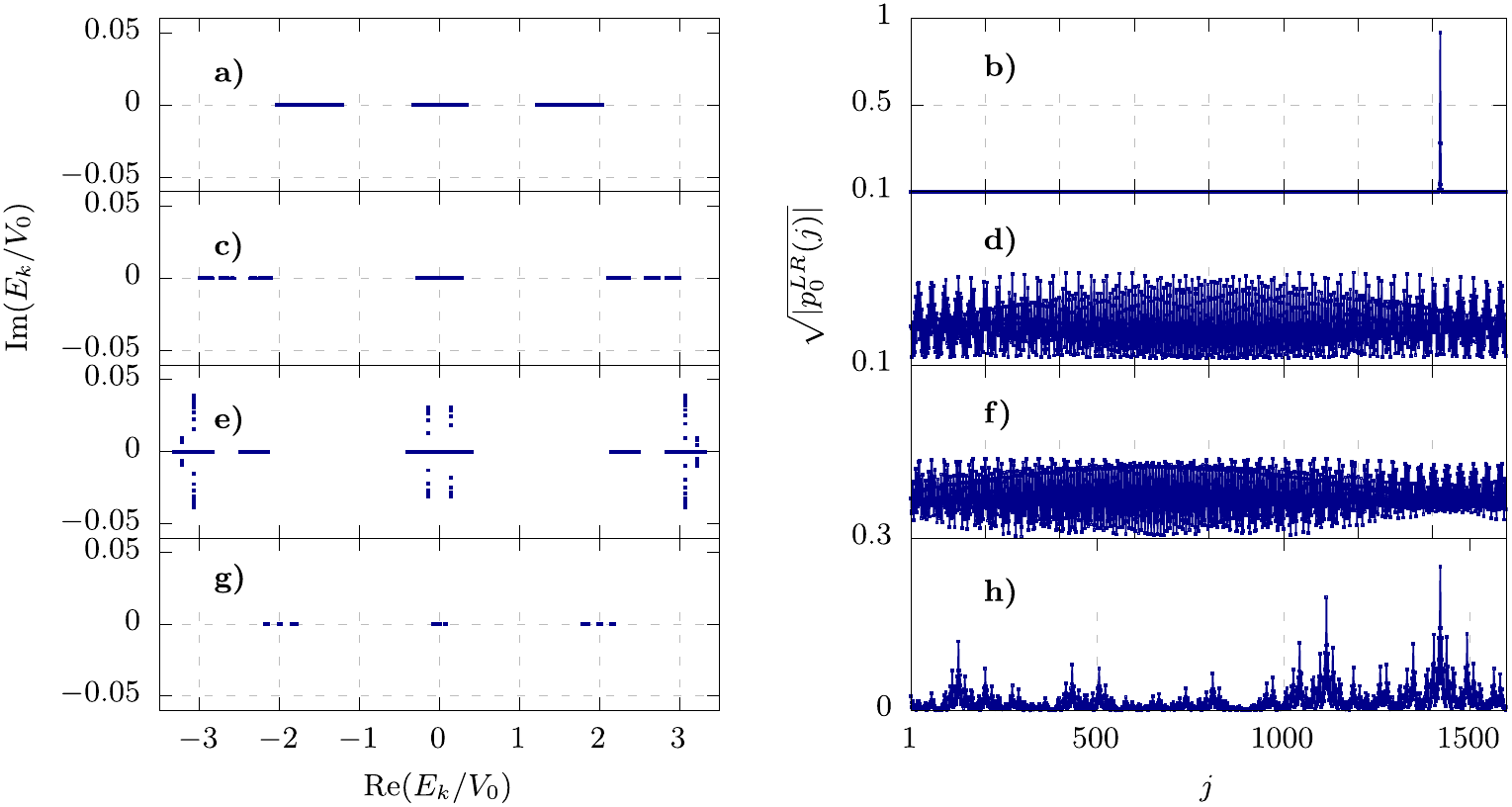}
	\caption{Energy spectra $\im (E_k)$ as a function of $\re (E_k)$ and $\sqrt {|p_0^{LR}(j)|}$ of the state with $E_0\approx0$ for $N=1597$ for the different sets of parameters marked in Fig. \ref{fig:Pic_PhaseDiagram} a). The spectra are shown in a) c) e) and g) and the wavefunction are shown in b) d) f) h) for P$_1$, P$_2$, P$_3$ and B$_{60}$ respectively.}\label{fig:Pic_Spectra}
\end{figure*}
We also perform an energy gap statistic analysis to diagnose the localization transition. As the energies can be complex in the $\mathcal{PT}$-broken regime, we restrict this analysis to the region $\gamma_0 \leq V_0$, where the averaged level spacing ratio is well defined: $r=\sum_k r_k / (N-1)$ and
\begin{align}
r_k  = \frac{\min(\delta_{k+1},\delta_{k})}{\max(\delta_{k+1},\delta_{k})}, \qquad\delta_{k}=E_{k+1}-E_k.
\end{align}
In the deeply localized region $R/V_0\rightarrow 0$, $\langle r \rangle \rightarrow \langle r \rangle_\textrm{Poisson}= 2 \ln(2)-1\approx0.3863$ for a Poisson distribution \cite{Oganesyan2007,Pal2010}, as shown in Fig \ref{fig:Pic_Localization} b). In the deep extended region $R/V_0 \rightarrow \infty$, an asymptotic degeneracy emerges due to the periodic boundary condition and vanishing disorder. Consequently, $\langle r \rangle \rightarrow 0$ in this limit, instead of $\langle r \rangle_\textrm{GOE} \approx 0.5307$ for a Gaussian orthogonal ensemble as one might na\"ively assume. As $\langle r \rangle$ also changes rapidly at $R=V_0$, this assures of a localization transition boundary as shown in Fig. \ref{fig:Pic_PhaseDiagram} b).

Our main results are summarized and illustrated in the phase diagrams in Fig. \ref{fig:Pic_PhaseDiagram}: (1) a robust $\mathcal{PT}$-symmetric phase exists for large system sizes and arbitrary disorder strength; (2) a disorder-driven localization transition occurs within the $\mathcal{PT}$-symmetric phase on a quarter circle arc $\sqrt{t^2+\gamma_0^2}=V_0$ as phase boundary; (3) along this phase boundary and $t=0,\gamma_0\geq V_0$, the system shows critical behavior; (4) in the $\mathcal{PT}$-broken phase the eigen wavefunctions are extended. 

\emph{Multifractal analysis.}--- Next, we investigate the spectra and wavefunctions at different phase regimes. As some typical examples, we show $E_k$ and $\sqrt{|p^{LR}_0(j)|}$ in Fig. \ref{fig:Pic_Spectra} for four sets of $\{t,\gamma_0\}$ marked in Fig. \ref{fig:Pic_PhaseDiagram} a) : $\rm P_1$ in the localized phase, $\rm P_2$ in the $\mathcal{PT}$-symmetric and extended phase, $\rm P_3$  in $\mathcal{PT}$-broken and extended phase, and $\rm B_{60}$ at the localization transition boundary. Here, $\sqrt{|p^{LR}_0(j)|}$ corresponds to the eigenstate with eigenenergy $E_0$ closest to $0$, which is near the center of the spectrum. The numerical examples are calculated using $\varphi \approx 0.157$ and $N=1597$. Figure \ref{fig:Pic_Spectra} a) and b) shows a purely real spectrum and localized wavefunction at $\rm P_1$. At $\rm P_2$, the spectrum is also purely real as shown in Fig. \ref{fig:Pic_Spectra} c), but the wavefunction spreads across all sites in Fig. \ref{fig:Pic_Spectra} d). Complex conjugate pairs show up in the spectrum in Fig. \ref{fig:Pic_Spectra} e), and the extended wavefunction is shown in Fig. \ref{fig:Pic_Spectra} f) for $\rm P_3$. In Fig. \ref{fig:Pic_Spectra}  g) and h), the spectrum and wavefunction for $R=V_0$ and $\theta=60^\circ$ ($\rm B_{60}$) are depicted. As this point is at the phase boundary between localized and extended region, we expect the system to show critical behavior. Indeed, looking at the wavefunction we can see that it is not completely smeared over the chain. The peaks are larger and the wavefunction looks less dense as for the extended states in Fig. \ref{fig:Pic_Spectra} d) and f). This is a signature of a multifractal wavefunction. To investigate the critical behavior of the system further, we employ a multifractal analysis.

To analyze the scaling behavior of the wavefunctions, we apply the approach detailed by Refs. \cite{Hiramoto1989, WangPRB2016, Supp} and only mention the key steps here. For a lattice with length $N=F_n$, where $F_n$ is the $n$-th Fibonacci number, a scaling index $\alpha_j$ can be defined as
\begin{equation}
|p_0^{\rm LR}(j)|=F_n^{-\alpha_j}.
\end{equation}
For an extended wavefunction, $\alpha_j \sim 1$ since $|p_0^{\rm LR}(j)|\sim1/F_n$. For a localized state, on the other hand, $|p_0^{\rm LR}(j)|$ is nonzero only on a finite number of lattice sites. Therefore, $\alpha_j \sim 0$ on these few localized sites and $\alpha_j \rightarrow \infty$ on the other sites. For critical wavefunctions, the index $\alpha_j$ would distribute on a finite interval $[\alpha_{\rm min}, \alpha_{\rm max}]$. Hence, we may use $\alpha_{\rm min}$ in the thermodynamic limit $n \rightarrow \infty$ to characterize the scaling behavior: $\alpha_{\rm min}=1$ for extended states, $\alpha_{\rm min}=0$ for localized states and $0<\alpha_{\rm min}<1$ for critical states. In the numerical calculations, we average $\alpha_{\rm min}$ over different quasi-disorder configurations for finite $n$. We fit the datapoints with a linear function to extrapolate the limit $1/n \rightarrow 0$. 

\begin{figure*}[t]
\includegraphics[scale=1]{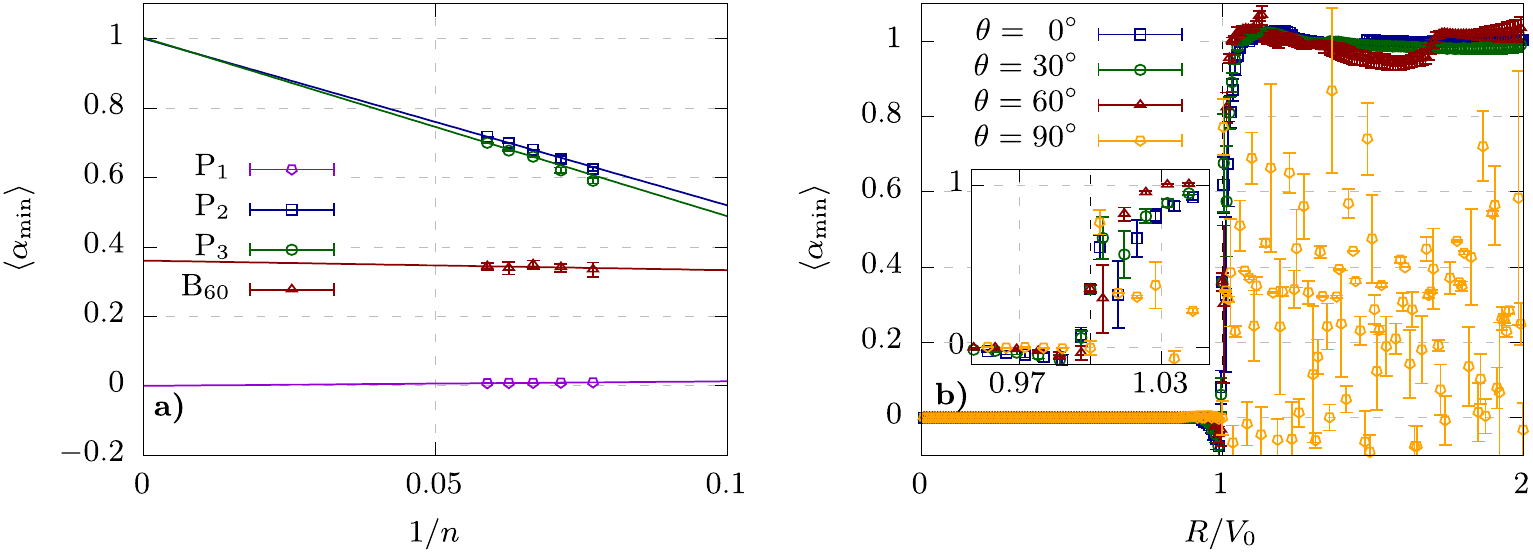}
\caption{a) shows $\langle \alpha_\textrm{min} \rangle$ for different chain length $N=F_n$ with $n=13-17$ for P$_1$, P$_2$, P$_3$ and B$_{60}$ defined in Fig. \ref{fig:Pic_PhaseDiagram} a). Extrapolation of $\langle \alpha_\textrm{min} \rangle$ to $1/n \rightarrow 0$ limit can distinguish extended, localized and critical phases. b) displays the values of $\langle \alpha_\textrm{min} \rangle$ for $1/n \rightarrow 0$ obtained from extrapolation for different $\theta$, illustrating the localization transition at $R=V_0$. The inset illustrate a zoom-in near $R=V_0$, emphasizing the critical index $\langle \alpha_\textrm{min} \rangle$ all collapse approximately on 0.36 for different $\theta$ except $\theta=90^\circ$.}\label{fig:Pic_Multifractal}
\end{figure*}

We present the results of the multifractal scaling in Fig. \ref{fig:Pic_Multifractal}. In Fig. \ref{fig:Pic_Multifractal} a), the purple pentagons correspond to $\rm P_1$ in the localized phase, where the extrapolation reveals $\langle \alpha_\textrm{min} \rangle \rightarrow 0$. Both the blue squares and green circles that correspond to $\rm P_2$ and $\rm P_3$ respectively show the trend $\langle \alpha_\textrm{min} \rangle \rightarrow 1$, confirming the wavefunctions are extended in both phases. At the localization transition boundary $\rm B_{60}$, the extrapolation of red triangles gives $\langle \alpha_\textrm{min} \rangle \approx0.361$, as a signature of the multifractal nature of the critical wavefunction. In Fig. \ref{fig:Pic_Multifractal} b) we display the extrapolated value of $\langle \alpha_\textrm{min} \rangle$ as a function of $R$. $\langle \alpha_\textrm{min} \rangle$ stays at zero for the localized phase region $R<V_0$. At the boundary, the value rises quickly in the critical region until the value assumes the extended one. At the critical point $R=V_0$ the value of $\langle \alpha_\textrm{min} \rangle\approx0.361\pm0.024$ stays constant for all simulated values of $\theta$ except $\theta=90^\circ$. The good agreement of $\langle \alpha_\textrm{min} \rangle$ between different $\theta$ at the critical point can be obvserved in the inset of Fig. \ref{fig:Pic_Multifractal} b) where we show the zoomed region around $R=V_0$, revealing the critical region within $R\in[0.96,1.04]$. We notice, $\theta=90^\circ$, $R>V_0$ correspond to the ``special ray'' mentioned earlier, where we don't average over $\varphi$, hence the finite-size effects become more severe. Nevertheless, despite the discontinuity and large error bars of $\alpha_{\rm min}$ on the ``special ray'', the wavefunction can be classified as multifractal as $0< \alpha_\textrm{min} <1$. This implies the system is critical at $t=0$ in the $\mathcal PT$-broken phase, which will be explored in a more systematic way in future studies.

\emph{Experimental realization.}--- Experimental realization of PT-symmetric Hamiltonian has been recently achieved in dissipative ultracold-atom systems via investigation of the dynamics conditioned on measurement outcomes \cite{LiNatComm2019, TakahashiPreprint2020}. Our model Hamiltonian Eq. (\ref{eqn:H}) can, in principle, be realized based on ultracold atoms in optical lattices with technologies in currently existing proposals such as engineered dissipation and laser-assisted hopping (see Supplemental Materials for details \cite{Supp}). 

\emph{Conclusion.}---We have studied a generalized $\mathcal{PT}$-symmetric AAH model. We have observed a $\mathcal{PT}$-symmetric phase $\gamma_0<V_0$ that is robust against disorder and system size. Furthermore, we have calculated the $\langle \textrm{MIPR}\rangle$ and carried out the energy gap statistics to charaterize the localized and extended phases. We report a localized phase within a quarter circle $\sqrt{\gamma_0^2+t^2}\leq V_0$. Additionally, the system features a critical behavior at the localization transition boundary $R=V_0$ and a special ray $\{R>V_0$, $\theta=90^\circ\}$, where we have analyzed fractal behaviors of the wavefunction.

\emph{Acknowledgement.}---We are grateful to Brendan C. Mulkerin for fruitful discussions. This research was supported by the Australian Research Council's (ARC) Discovery Program, Grant No. DP180102018 (X.-J.L), Grant No. DP170104008 (H.H.), and Grants No. DE180100592 and No. DP190100815 (J.W.).

\appendix

\section*{Supplemental Material for Anderson localization transition in a  robust $\mathcal{PT}$-symmetric phase of a generalized Aubry–André model}

\section{Symmetries of the system}
\begin{figure}[htpb]
\includegraphics[scale=1]{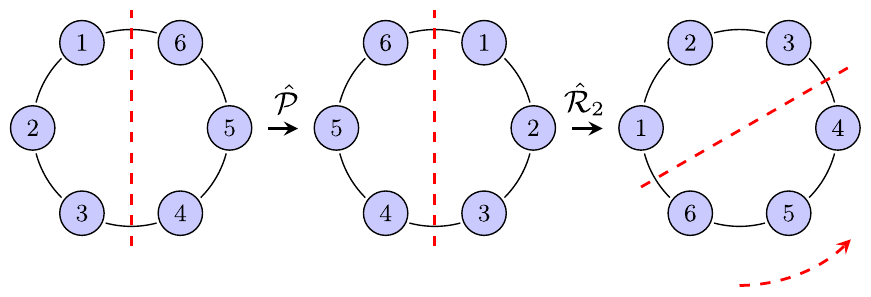}
\caption{An illustration of the  $\hat{\mc{P}}$ and $\hat {\mathcal R}_k$ operators on a $N=6$ lattices subject to periodic boundary condition.}\label{sketch}
\end{figure}

$\mathcal{PT}$-symmetry refers to a combined parity $\mathcal P$ and time-reversal $\mathcal T$ symmetry. The effects of corresponding space-reflection operator $\mathcal{\hat P}$ and time-reversal operator $\mathcal{\hat T}$ on a discrete system are,
\begin{equation}
\mathcal{\hat T} i \mathcal{\hat T}=-i,\ \ \ \mathcal{\hat P} \hat c_{j}^{\dagger} \mathcal{\hat P}=\hat c_{N+1-j}^{\dagger}.
\end{equation}
Applying the combined $\mathcal {\hat P \hat T}$ operator to our model Hamiltonian Eq. (1) in the main text yields
\begin{equation}
\mathcal{\hat P \hat T} \hat H \left(\varphi\right) \mathcal{\hat T \hat P}=\hat H \left(\bar \varphi\right),
\end{equation}
where $\bar \varphi = -2\pi\beta (N+1) - \varphi \pmod{2\pi}$. We have applied $\sin (- \phi)=-\sin \phi$ and $\cos (-\phi)=\cos \phi$ in the derivation. Here, we adopt $\beta=M/N$, where $M$ and $N$ are mutually prime. Therefore, one can verify that if $\varphi=\pi-\pi\beta \pmod{2\pi}$ or $2\pi-\pi\beta \pmod{2\pi}$, $\bar \varphi = \varphi \pmod{2\pi}$, i.e., the Hamiltonian is $\mathcal {PT}$-symmetric. 
\begin{figure*}[t]
\includegraphics[scale=1]{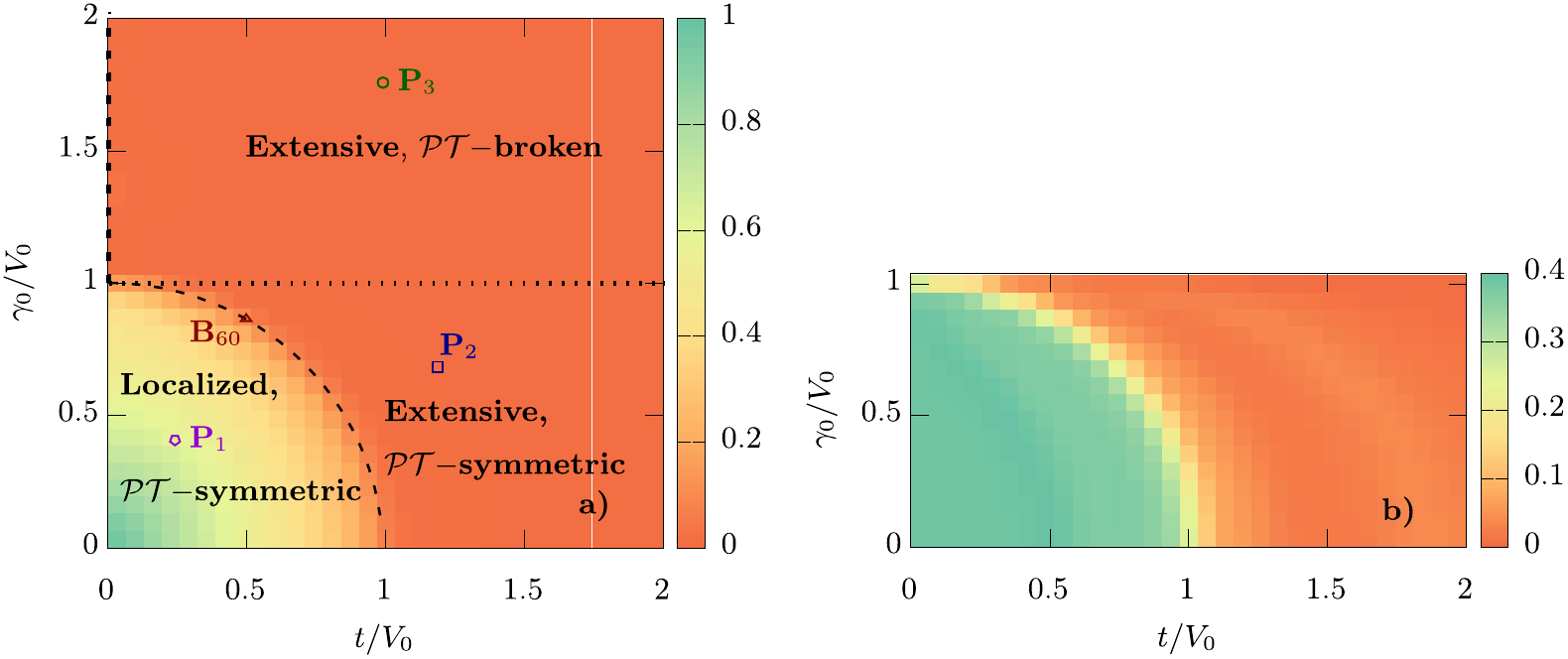}    
    	\caption{Phase diagrams for the MIPR in subfig. a) and $\langle r\rangle$ ins subfig. b) of the system for $N=233$ at infinite temperature. The MIPR in the localized region in subfig. a) has decreased but is still clearly visible. The gap statistics in subfig. b) seems to be almost indistinguishable from Fig. 1 in the main text. No trace of a mobility edge was found in the system.}
		\label{fig:Pic_5.7}
\end{figure*}
A key feature of  $\mathcal {PT}$-symmetric Hamiltonian is its purely real or complex-conjugate-pairs spectrum. Nevertheless, in principle, such features only requires the existence of an antiunitary operator to commute with the Hamiltonian. We can construct an antiunitary operator as $\hat {\mathcal  A}_k = \hat {\mathcal R}_k \hat {\mathcal P} \hat{ \mathcal T}$, where $\mathcal R_k$ is a unitary operator ``rotating'' the system by $k\in\mathbb{Z}$ sites in the counter-clockwise direction (see Fig. \ref{sketch} for an illustrative example):
\begin{equation}
\mathcal R _k^\dagger c_j^\dagger \mathcal R_k=c_{j+k}^\dagger.
\end{equation} 
Applying $\mathcal R _k$ on the Hamiltonian with periodic boundary conditions gives
\begin{equation}
\mathcal{\hat R}_k^\dagger \hat H \left(\varphi\right) \mathcal{\hat R}_k=\hat H \left(\varphi_k \right),
\end{equation}
where $\varphi_k=\varphi-2\pi \beta k$. Since $\mathcal{\hat R}_k$ is unitary, the spectra of  $\hat H \left(\varphi \right)$ and $\hat H \left(\varphi_k\right)$ are the same. In addition, from the theorem of modular inverses \cite{Rosen2005}, there is a solution to $M k_1 =1 \pmod{N}$ if and only if $M$ and $N$ are coprime, which implies we can always find a $k_1=M^{-1} \pmod{N}$ that satisfies $\varphi_{k_1}=\varphi-2 \pi/N$. Therefore, the spectrum of $\hat H \left(\varphi \right)$ is periodic as a function of $\varphi$ with periodicity $2\pi/N$. As a result, the spectrum of $\hat H \left(\varphi \right)$ are always real or complex-conjutate-pairs, i.e. $\mathcal {PT}$-symmetric, for $\varphi = \pi-\beta\pi+2 k \pi/N \pmod{2\pi}$ or $2\pi-\beta\pi+2 k \pi/N \pmod{2\pi}$. This condition is equivalent to $\varphi=(2k+1)\pi/N \pmod{2\pi}$ for even chains and $\varphi=k\pi/N \pmod{2\pi}$ for odd chains with $k\in\mathbb{Z}$. The number of these ``$\mathcal {PT}$-symmetric'' points of $\varphi$ becomes infinite for $N \rightarrow \infty$, and the spacing between adjacent points vanishes. The periodicity of the spectrum also gives a technical benefit: we only need to average $\varphi$ over $[0,2\pi/N)$ to emulate the disorder realisation average.  

\section{Phase diagram for $E\approx0$}
In order to rule out the possibility of a mobility edge we simulate $\langle \textrm{MIPR}\rangle $ and $\langle r\rangle$ . We calculate those observables for the 50 eigenstates with their real part of the energy closest to $E\approx0$ which corresonds to the infinite temperature limit. In Fig. \ref{fig:Pic_5.7} the phase diagrams are depicted. The $\langle \textrm{MIPR}\rangle $ in subfig. a) in the semicircle $R\leq V_0$ has decreased in value, but is clearly non-zero. The gap statistics in subfig. b) is indestinguishable from Fig. 1 b). We conclude that a localized region still exists at $T\rightarrow\infty$ and no trace of a mobility edge was found in the system.

\section{Multifractal}
For a lattice of size $F_n$ where $F_n$ is the $n$-th Fibonacci number, $p_k^{\rm LR}(j) = |\psi_k^L(j)\psi_k^R(j)|$,  $j\in [1,F_n]$ plays the role of the onsite probabilities for a given eigenstate with energy $E_k$. We usually select the energy $E_0$ closest to $E=0$. Depending on whether the system is in an extensive phase or localized phase, the wavefunction either can be smeared over the lattice or be localized at a single site. Generally, to allow a smooth transition between the two cases we can define a scaling exponent $\alpha$ such that
\begin{equation}
p^{(n)}_0(j)=F_n^{-\alpha_j^{(n)}}.
\end{equation}
At the localization phase transition boundary, $\alpha_j^{(n)}$ distribute between $[\alpha^{(n)}_\textrm{min} ,\alpha^{(n)}_\textrm{max}]$. The multifractal analysis for a given level of approximation $n$ can be extrapolated in the thermodynamic limit as $\alpha_\textrm{min}=\lim_{n\rightarrow\infty}\alpha^{(n)}_\textrm{min}$. Thus by identifying the minimum value $\alpha_\textrm{min}$ we can classify the wavefunction as extended for $\alpha_\textrm{min}=1$, critical for $1< \alpha_\textrm{min}< 0$ and localized for $\alpha_\textrm{min}=0$.  For the numerical calculation of $\alpha_\textrm{min}$, we follow the work by Hiramoto and Kohmoto \cite{Hiramoto1989, WangPRB2016} who treat $\alpha_\textrm{min}$ as energy of a canonical system. They then define an entropy that can be related to the onsite probabilities of the chain. The scaling exponent is given in terms of a parameter $q\in \mathbb{R}$
\begin{align}
\alpha=\frac{-1}{n\epsilon}\frac{\textrm{d}\ln[Z_n(q)]}{\textrm{d}q},\qquad Z_n(q)=\sum_{j=1}^{F_n}p_j^q
\end{align}
where $\epsilon=\ln[( \sqrt{5} + 1)/2]$ is the logarithm of the golden ratio. We vary $q$ to find the minimum of $\alpha$ for chain lengths between $N=233-1597$. Shorter chains were omitted as $n<13$ was not sufficient for the scaling. For the numerical calculations we average over the quasi-disorder configurations and we fit the datapoints with a linear function to extrapolate the $1/n \rightarrow 0$ limit.

\section{Experimental realization.}
\begin{figure}[htpb]
\includegraphics[width=0.98 \columnwidth]{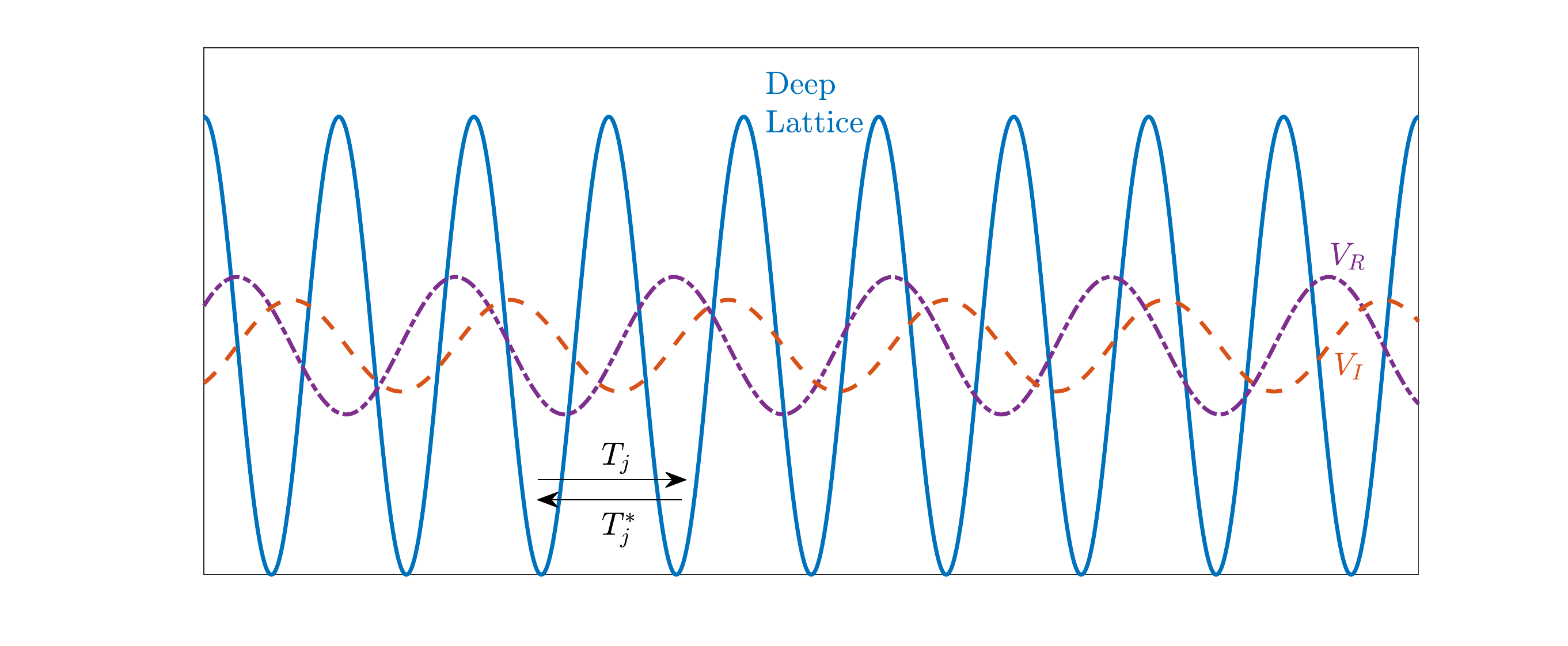}
\caption{A sketch of a proposed experimental realization. }\label{ExpRealization}
\end{figure}

We here show that it is possible to realize a non-Hermitian system in an ultracold atomic system with similar technologies proposed in previous studies \cite{BlochPRL2011, SpielmanPRL2012, LiNatComm2019, TakahashiPreprint2020}. The model Hamiltonian, Eq. (1) in the main text, can be written as
\begin{equation}
\hat H = \sum_{j=1}^N \left[ t_j^R \hat{c}_{j+1}^{\dagger}\hat{c}_{j}+ t_j^L \hat{c}_{j}^{\dagger}\hat{c}_{j+1} + V_j \hat{c}_{j}^{\dagger}\hat{c}_{j}\right],
\end{equation}
where $t_j^R\equiv t_j=t+\ii\gamma_{0}\sin(2\pi\beta j+\varphi)$, $t_j^L\equiv t_{j+1}=t+\ii\gamma_{0}\sin(2\pi\beta j+2\pi\beta \varphi)$ and $V_j=2V_{0}\cos(2\pi\beta j+\varphi)$. The hopping parameters can be expressed as
\begin{equation}
t_j^R = T_j+i\Gamma_j, t_j^L= T_j^*+\ii\Gamma_j.
\end{equation}
for convenience. Here, $T_j=(t_j+t_{j+1}^*)/2$ and $\Gamma_j=(t_j- t_{j+1}^*)/2 \ii$ represent the Hermitian and anti-Hermitian hopping amplitude, respectively. A manipulation of algebra gives,
\begin{equation}
T_j=t- \ii \gamma_0 \sin(\pi \beta) \cos [2\pi\beta (j+1/2) +\varphi],
\end{equation}
and
\begin{equation}
\Gamma_j= \gamma_0 \cos (\pi \beta) \sin [2\pi\beta (j+1/2) +\varphi].
\end{equation}
 The Hamiltonian can therefore be written as $\hat H = \hat K_H + \hat K_A +\hat V$, where
\begin{equation}
\hat K_H = \sum_{j=1}^N \left[ T_j \hat{c}_{j+1}^{\dagger}\hat{c}_{j}+ {\mathrm {h. c.}} \right],
\end{equation}
\begin{equation}
\hat K_A = \sum_{j=1}^N \ii \Gamma_j \left[ \hat{c}_{j+1}^{\dagger}\hat{c}_{j}+ \hat{c}_{j}^{\dagger}\hat{c}_{j+1} \right],
\end{equation}
and
\begin{equation}
\hat V = \sum_{j=1}^N V_j \hat{c}_{j}^{\dagger}\hat{c}_{j}.
\end{equation}
The Hermitian but complex hopping term $\hat K_H$ can be realized via laser-assisted hopping, where the complex phase is associated with the laser photon's momentum \cite{BlochPRL2011, SpielmanPRL2012}. The on-site potential $\hat V$ and anti-Hermitian hopping $\hat K_A$ can be realized via a pair of far-detuned and weak near-resonant standing waves that have different wave-length from a deep lattice as indicated in Fig. (\ref{ExpRealization}). It has been shown in Ref. \cite{YutoNC2017} that the effects of the far-detuned and weak near-resonant standing wave can be regarded as introducing a real potential $V_R$ and an imaginary one $V_I$ respectively,
\begin{equation}
V_R=U_R\cos(2\pi \beta x /a +\varphi)
\end{equation}
\begin{equation}
V_I=\ii U_I\sin(2\pi \beta x /a +\varphi)
\end{equation}
where $a$ is the lattice constant. In a tight-biding approximation, $V_R$ gives the on-site modulation $\sum_{j=1}^N V_j \hat{c}_{j}^{\dagger}\hat{c}_{j}$, where
\begin{equation}
V_j=\int dx W_j(x) U_R\cos(2\pi \beta x /a +\varphi) W_j(x) \propto \cos(2\pi \beta j +\varphi),
\end{equation}
where $W_j(x)$ are Wannier mode that localized at site $j$. Similarly,  $V_I$ gives an anti-Hermitian hopping:
\begin{eqnarray}
\Gamma_j &= & \int dx W_j(x) U_I\cos(2\pi \beta x /a +\varphi) W_{j+1}(x) \\ \nonumber
& \propto & \sin[2\pi \beta (j+1/2) +\varphi],
\end{eqnarray}
where $W_j(x) W_{j+1}(x)$ has a maximum in the middle of site $j$ and $j+1$. $V_I$ will also introduce an on-site loss, which can be neglected via the renormalization \cite{LiNatComm2019} or postselection procedure \cite{YutoNC2017, GongPRX2018}. Therefore, by controlling of $U_R$ and $U_I$, we can tune the parameters to realize our model Hamiltonian.

\bibliographystyle{apsrev4-1}
\bibliography{references}

\begin{thebibliography}{82}%
\makeatletter
\providecommand \@ifxundefined [1]{%
 \@ifx{#1\undefined}
}%
\providecommand \@ifnum [1]{%
 \ifnum #1\expandafter \@firstoftwo
 \else \expandafter \@secondoftwo
 \fi
}%
\providecommand \@ifx [1]{%
 \ifx #1\expandafter \@firstoftwo
 \else \expandafter \@secondoftwo
 \fi
}%
\providecommand \natexlab [1]{#1}%
\providecommand \enquote  [1]{``#1''}%
\providecommand \bibnamefont  [1]{#1}%
\providecommand \bibfnamefont [1]{#1}%
\providecommand \citenamefont [1]{#1}%
\providecommand \href@noop [0]{\@secondoftwo}%
\providecommand \href [0]{\begingroup \@sanitize@url \@href}%
\providecommand \@href[1]{\@@startlink{#1}\@@href}%
\providecommand \@@href[1]{\endgroup#1\@@endlink}%
\providecommand \@sanitize@url [0]{\catcode `\\12\catcode `\$12\catcode
  `\&12\catcode `\#12\catcode `\^12\catcode `\_12\catcode `\%12\relax}%
\providecommand \@@startlink[1]{}%
\providecommand \@@endlink[0]{}%
\providecommand \url  [0]{\begingroup\@sanitize@url \@url }%
\providecommand \@url [1]{\endgroup\@href {#1}{\urlprefix }}%
\providecommand \urlprefix  [0]{URL }%
\providecommand \Eprint [0]{\href }%
\providecommand \doibase [0]{http://dx.doi.org/}%
\providecommand \selectlanguage [0]{\@gobble}%
\providecommand \bibinfo  [0]{\@secondoftwo}%
\providecommand \bibfield  [0]{\@secondoftwo}%
\providecommand \translation [1]{[#1]}%
\providecommand \BibitemOpen [0]{}%
\providecommand \bibitemStop [0]{}%
\providecommand \bibitemNoStop [0]{.\EOS\space}%
\providecommand \EOS [0]{\spacefactor3000\relax}%
\providecommand \BibitemShut  [1]{\csname bibitem#1\endcsname}%
\let\auto@bib@innerbib\@empty
%</preamble>
\bibitem [{\citenamefont {Sokolov}\ and\ \citenamefont
  {Zelevinsky}(1988)}]{SololovPLB1988}%
  \BibitemOpen
  \bibfield  {author} {\bibinfo {author} {\bibfnamefont {V.~V.}\ \bibnamefont
  {Sokolov}}\ and\ \bibinfo {author} {\bibfnamefont {V.~G.}\ \bibnamefont
  {Zelevinsky}},\ }\href@noop {} {\bibfield  {journal} {\bibinfo  {journal}
  {Phys. Lett. B}\ }\textbf {\bibinfo {volume} {202}},\ \bibinfo {pages} {10}
  (\bibinfo {year} {1988})}\BibitemShut {NoStop}%
\bibitem [{\citenamefont {Sokolov}\ and\ \citenamefont
  {Zelevinsky}(1989)}]{SololovNPA1989}%
  \BibitemOpen
  \bibfield  {author} {\bibinfo {author} {\bibfnamefont {V.~V.}\ \bibnamefont
  {Sokolov}}\ and\ \bibinfo {author} {\bibfnamefont {V.~G.}\ \bibnamefont
  {Zelevinsky}},\ }\href@noop {} {\bibfield  {journal} {\bibinfo  {journal}
  {Nucl. Phys. A}\ }\textbf {\bibinfo {volume} {504}},\ \bibinfo {pages} {562}
  (\bibinfo {year} {1989})}\BibitemShut {NoStop}%
\bibitem [{\citenamefont {Rotter}(1991)}]{RotterRPP1991}%
  \BibitemOpen
  \bibfield  {author} {\bibinfo {author} {\bibfnamefont {I.}~\bibnamefont
  {Rotter}},\ }\href@noop {} {\bibfield  {journal} {\bibinfo  {journal} {Rep.
  Prog. Phys.}\ }\textbf {\bibinfo {volume} {54}},\ \bibinfo {pages} {635}
  (\bibinfo {year} {1991})}\BibitemShut {NoStop}%
\bibitem [{\citenamefont {Sokolov}\ and\ \citenamefont
  {Zelevinsky}(1992)}]{SololovAP1992}%
  \BibitemOpen
  \bibfield  {author} {\bibinfo {author} {\bibfnamefont {V.~V.}\ \bibnamefont
  {Sokolov}}\ and\ \bibinfo {author} {\bibfnamefont {V.~G.}\ \bibnamefont
  {Zelevinsky}},\ }\href@noop {} {\bibfield  {journal} {\bibinfo  {journal}
  {Ann. Phys. (NY)}\ }\textbf {\bibinfo {volume} {216}},\ \bibinfo {pages}
  {323} (\bibinfo {year} {1992})}\BibitemShut {NoStop}%
\bibitem [{\citenamefont {Carmichael}(1993)}]{Carmichael1993}%
  \BibitemOpen
  \bibfield  {author} {\bibinfo {author} {\bibfnamefont {H.}~\bibnamefont
  {Carmichael}},\ }\href@noop {} {\emph {\bibinfo {title} {An Open System
  Approach to Quantum Optics}}}\ (\bibinfo  {publisher} {Springer,},\ \bibinfo
  {year} {1993})\BibitemShut {NoStop}%
\bibitem [{\citenamefont {Dittes}(2000)}]{DittesPR2000}%
  \BibitemOpen
  \bibfield  {author} {\bibinfo {author} {\bibfnamefont {F.~M.}\ \bibnamefont
  {Dittes}},\ }\href@noop {} {\bibfield  {journal} {\bibinfo  {journal} {Phys.
  Rep.}\ }\textbf {\bibinfo {volume} {339}},\ \bibinfo {pages} {215} (\bibinfo
  {year} {2000})}\BibitemShut {NoStop}%
\bibitem [{\citenamefont {Daley}(2014)}]{DaleyAP2014}%
  \BibitemOpen
  \bibfield  {author} {\bibinfo {author} {\bibfnamefont {A.~J.}\ \bibnamefont
  {Daley}},\ }\href@noop {} {\bibfield  {journal} {\bibinfo  {journal} {Adv.
  Phys.}\ }\textbf {\bibinfo {volume} {63}},\ \bibinfo {pages} {77} (\bibinfo
  {year} {2014})}\BibitemShut {NoStop}%
\bibitem [{\citenamefont {Ashida}\ \emph {et~al.}(2016)\citenamefont {Ashida},
  \citenamefont {Furukawa},\ and\ \citenamefont {Ueda}}]{YutoPRA2016}%
  \BibitemOpen
  \bibfield  {author} {\bibinfo {author} {\bibfnamefont {Y.}~\bibnamefont
  {Ashida}}, \bibinfo {author} {\bibfnamefont {S.}~\bibnamefont {Furukawa}}, \
  and\ \bibinfo {author} {\bibfnamefont {M.}~\bibnamefont {Ueda}},\ }\href@noop
  {} {\bibfield  {journal} {\bibinfo  {journal} {Phys. Rev. A}\ }\textbf
  {\bibinfo {volume} {94}},\ \bibinfo {pages} {053615} (\bibinfo {year}
  {2016})}\BibitemShut {NoStop}%
\bibitem [{\citenamefont {Ashida}\ \emph {et~al.}(2017)\citenamefont {Ashida},
  \citenamefont {Furukawa},\ and\ \citenamefont {Ueda}}]{YutoNC2017}%
  \BibitemOpen
  \bibfield  {author} {\bibinfo {author} {\bibfnamefont {Y.}~\bibnamefont
  {Ashida}}, \bibinfo {author} {\bibfnamefont {S.}~\bibnamefont {Furukawa}}, \
  and\ \bibinfo {author} {\bibfnamefont {M.}~\bibnamefont {Ueda}},\ }\href@noop
  {} {\bibfield  {journal} {\bibinfo  {journal} {Nat. Comm.}\ }\textbf
  {\bibinfo {volume} {8}},\ \bibinfo {pages} {15791} (\bibinfo {year}
  {2017})}\BibitemShut {NoStop}%
\bibitem [{\citenamefont {Jin}\ and\ \citenamefont {Song}(2019)}]{JinPRB2019}%
  \BibitemOpen
  \bibfield  {author} {\bibinfo {author} {\bibfnamefont {L.}~\bibnamefont
  {Jin}}\ and\ \bibinfo {author} {\bibfnamefont {Z.}~\bibnamefont {Song}},\
  }\href@noop {} {\bibfield  {journal} {\bibinfo  {journal} {Phys. Rev. B}\
  }\textbf {\bibinfo {volume} {99}},\ \bibinfo {pages} {081103} (\bibinfo
  {year} {2019})}\BibitemShut {NoStop}%
\bibitem [{\citenamefont {Bender}\ and\ \citenamefont
  {Boettcher}(1998)}]{BenderPRL1998}%
  \BibitemOpen
  \bibfield  {author} {\bibinfo {author} {\bibfnamefont {C.~M.}\ \bibnamefont
  {Bender}}\ and\ \bibinfo {author} {\bibfnamefont {S.}~\bibnamefont
  {Boettcher}},\ }\href@noop {} {\bibfield  {journal} {\bibinfo  {journal}
  {Phys. Rev. Lett.}\ }\textbf {\bibinfo {volume} {80}},\ \bibinfo {pages}
  {5243} (\bibinfo {year} {1998})}\BibitemShut {NoStop}%
\bibitem [{\citenamefont {Bender}\ \emph {et~al.}(1999)\citenamefont {Bender},
  \citenamefont {Boettcher},\ and\ \citenamefont {Meisinger}}]{BenderJMP1999}%
  \BibitemOpen
  \bibfield  {author} {\bibinfo {author} {\bibfnamefont {C.~M.}\ \bibnamefont
  {Bender}}, \bibinfo {author} {\bibfnamefont {S.}~\bibnamefont {Boettcher}}, \
  and\ \bibinfo {author} {\bibfnamefont {P.~N.}\ \bibnamefont {Meisinger}},\
  }\href@noop {} {\bibfield  {journal} {\bibinfo  {journal} {J. Math. Phys.
  (NY)}\ }\textbf {\bibinfo {volume} {40}},\ \bibinfo {pages} {2201} (\bibinfo
  {year} {1999})}\BibitemShut {NoStop}%
\bibitem [{\citenamefont {Bender}\ \emph {et~al.}(2002)\citenamefont {Bender},
  \citenamefont {Brody},\ and\ \citenamefont {Jones}}]{BenderPRL2002}%
  \BibitemOpen
  \bibfield  {author} {\bibinfo {author} {\bibfnamefont {C.~M.}\ \bibnamefont
  {Bender}}, \bibinfo {author} {\bibfnamefont {D.~C.}\ \bibnamefont {Brody}}, \
  and\ \bibinfo {author} {\bibfnamefont {H.~F.}\ \bibnamefont {Jones}},\
  }\href@noop {} {\bibfield  {journal} {\bibinfo  {journal} {Phys. Rev. Lett.}\
  }\textbf {\bibinfo {volume} {89}},\ \bibinfo {pages} {270401} (\bibinfo
  {year} {2002})}\BibitemShut {NoStop}%
\bibitem [{\citenamefont {Bender}\ \emph {et~al.}(2007)\citenamefont {Bender},
  \citenamefont {Boettcher},\ and\ \citenamefont {Meisinger}}]{BenderRPP2007}%
  \BibitemOpen
  \bibfield  {author} {\bibinfo {author} {\bibfnamefont {C.~M.}\ \bibnamefont
  {Bender}}, \bibinfo {author} {\bibfnamefont {S.}~\bibnamefont {Boettcher}}, \
  and\ \bibinfo {author} {\bibfnamefont {P.~N.}\ \bibnamefont {Meisinger}},\
  }\href@noop {} {\bibfield  {journal} {\bibinfo  {journal} {Rep. Prog. Phys.}\
  }\textbf {\bibinfo {volume} {70}},\ \bibinfo {pages} {947} (\bibinfo {year}
  {2007})}\BibitemShut {NoStop}%
\bibitem [{\citenamefont {Bender}\ \emph {et~al.}(2017)\citenamefont {Bender},
  \citenamefont {Brody},\ and\ \citenamefont {M\"uller}}]{BenderPRL2017}%
  \BibitemOpen
  \bibfield  {author} {\bibinfo {author} {\bibfnamefont {C.~M.}\ \bibnamefont
  {Bender}}, \bibinfo {author} {\bibfnamefont {D.~C.}\ \bibnamefont {Brody}}, \
  and\ \bibinfo {author} {\bibfnamefont {M.~P.}\ \bibnamefont {M\"uller}},\
  }\href@noop {} {\bibfield  {journal} {\bibinfo  {journal} {Phys. Rev. Lett.}\
  }\textbf {\bibinfo {volume} {118}},\ \bibinfo {pages} {130201} (\bibinfo
  {year} {2017})}\BibitemShut {NoStop}%
\bibitem [{\citenamefont {Bendix}\ \emph {et~al.}(2009)\citenamefont {Bendix},
  \citenamefont {Fleischmann}, \citenamefont {Kottos},\ and\ \citenamefont
  {Shapiro}}]{BendixPRL2009}%
  \BibitemOpen
  \bibfield  {author} {\bibinfo {author} {\bibfnamefont {O.}~\bibnamefont
  {Bendix}}, \bibinfo {author} {\bibfnamefont {R.}~\bibnamefont {Fleischmann}},
  \bibinfo {author} {\bibfnamefont {T.}~\bibnamefont {Kottos}}, \ and\ \bibinfo
  {author} {\bibfnamefont {B.}~\bibnamefont {Shapiro}},\ }\href@noop {}
  {\bibfield  {journal} {\bibinfo  {journal} {Phys. Rev. Lett.}\ }\textbf
  {\bibinfo {volume} {103}},\ \bibinfo {pages} {030402} (\bibinfo {year}
  {2009})}\BibitemShut {NoStop}%
\bibitem [{\citenamefont {Jin}\ and\ \citenamefont {Song}(2009)}]{JinPRA2009}%
  \BibitemOpen
  \bibfield  {author} {\bibinfo {author} {\bibfnamefont {L.}~\bibnamefont
  {Jin}}\ and\ \bibinfo {author} {\bibfnamefont {Z.}~\bibnamefont {Song}},\
  }\href@noop {} {\bibfield  {journal} {\bibinfo  {journal} {Phys. Rev. A}\
  }\textbf {\bibinfo {volume} {80}},\ \bibinfo {pages} {052107} (\bibinfo
  {year} {2009})}\BibitemShut {NoStop}%
\bibitem [{\citenamefont {El-Ganainy}\ \emph {et~al.}(2007)\citenamefont
  {El-Ganainy}, \citenamefont {Makris}, \citenamefont {Christodoulides},\ and\
  \citenamefont {Musslimani}}]{El-GanainyOL2007}%
  \BibitemOpen
  \bibfield  {author} {\bibinfo {author} {\bibfnamefont {R.}~\bibnamefont
  {El-Ganainy}}, \bibinfo {author} {\bibfnamefont {K.~G.}\ \bibnamefont
  {Makris}}, \bibinfo {author} {\bibfnamefont {D.~N.}\ \bibnamefont
  {Christodoulides}}, \ and\ \bibinfo {author} {\bibfnamefont {Z.~H.}\
  \bibnamefont {Musslimani}},\ }\href@noop {} {\bibfield  {journal} {\bibinfo
  {journal} {Opt. Lett.}\ }\textbf {\bibinfo {volume} {32}},\ \bibinfo {pages}
  {2632} (\bibinfo {year} {2007})}\BibitemShut {NoStop}%
\bibitem [{\citenamefont {Makris}\ \emph {et~al.}(2008)\citenamefont {Makris},
  \citenamefont {El-Ganainy}, \citenamefont {Christodoulides},\ and\
  \citenamefont {Musslimani}}]{MakrisPRL2008}%
  \BibitemOpen
  \bibfield  {author} {\bibinfo {author} {\bibfnamefont {K.~G.}\ \bibnamefont
  {Makris}}, \bibinfo {author} {\bibfnamefont {R.}~\bibnamefont {El-Ganainy}},
  \bibinfo {author} {\bibfnamefont {D.~N.}\ \bibnamefont {Christodoulides}}, \
  and\ \bibinfo {author} {\bibfnamefont {Z.~H.}\ \bibnamefont {Musslimani}},\
  }\href@noop {} {\bibfield  {journal} {\bibinfo  {journal} {Phys. Rev. Lett.}\
  }\textbf {\bibinfo {volume} {100}},\ \bibinfo {pages} {103904} (\bibinfo
  {year} {2008})}\BibitemShut {NoStop}%
\bibitem [{\citenamefont {Musslimani}\ \emph {et~al.}(2008)\citenamefont
  {Musslimani}, \citenamefont {Makris}, \citenamefont {El-Ganainy},\ and\
  \citenamefont {Christodoulides}}]{MusslimaniPRL2008}%
  \BibitemOpen
  \bibfield  {author} {\bibinfo {author} {\bibfnamefont {Z.~H.}\ \bibnamefont
  {Musslimani}}, \bibinfo {author} {\bibfnamefont {K.~G.}\ \bibnamefont
  {Makris}}, \bibinfo {author} {\bibfnamefont {R.}~\bibnamefont {El-Ganainy}},
  \ and\ \bibinfo {author} {\bibfnamefont {D.~N.}\ \bibnamefont
  {Christodoulides}},\ }\href@noop {} {\bibfield  {journal} {\bibinfo
  {journal} {Phys. Rev. Lett.}\ }\textbf {\bibinfo {volume} {100}},\ \bibinfo
  {pages} {030402} (\bibinfo {year} {2008})}\BibitemShut {NoStop}%
\bibitem [{\citenamefont {Longhi}(2009{\natexlab{a}})}]{LonghiPRL2009}%
  \BibitemOpen
  \bibfield  {author} {\bibinfo {author} {\bibfnamefont {S.}~\bibnamefont
  {Longhi}},\ }\href@noop {} {\bibfield  {journal} {\bibinfo  {journal} {Phys.
  Rev. Lett.}\ }\textbf {\bibinfo {volume} {103}},\ \bibinfo {pages} {123601}
  (\bibinfo {year} {2009}{\natexlab{a}})}\BibitemShut {NoStop}%
\bibitem [{\citenamefont {Longhi}(2009{\natexlab{b}})}]{LonghiPRB2009}%
  \BibitemOpen
  \bibfield  {author} {\bibinfo {author} {\bibfnamefont {S.}~\bibnamefont
  {Longhi}},\ }\href@noop {} {\bibfield  {journal} {\bibinfo  {journal} {Phys.
  Rev. B}\ }\textbf {\bibinfo {volume} {80}},\ \bibinfo {pages} {235102}
  (\bibinfo {year} {2009}{\natexlab{b}})}\BibitemShut {NoStop}%
\bibitem [{\citenamefont {Guo}\ \emph {et~al.}(2009)\citenamefont {Guo},
  \citenamefont {Salamo}, \citenamefont {Duchesne}, \citenamefont {Morandotti},
  \citenamefont {Volatier-Ravat}, \citenamefont {Aimez}, \citenamefont
  {Siviloglou},\ and\ \citenamefont {Christodoulides}}]{GuoPRL2009}%
  \BibitemOpen
  \bibfield  {author} {\bibinfo {author} {\bibfnamefont {A.}~\bibnamefont
  {Guo}}, \bibinfo {author} {\bibfnamefont {G.~J.}\ \bibnamefont {Salamo}},
  \bibinfo {author} {\bibfnamefont {D.}~\bibnamefont {Duchesne}}, \bibinfo
  {author} {\bibfnamefont {R.}~\bibnamefont {Morandotti}}, \bibinfo {author}
  {\bibfnamefont {M.}~\bibnamefont {Volatier-Ravat}}, \bibinfo {author}
  {\bibfnamefont {V.}~\bibnamefont {Aimez}}, \bibinfo {author} {\bibfnamefont
  {G.~A.}\ \bibnamefont {Siviloglou}}, \ and\ \bibinfo {author} {\bibfnamefont
  {D.~N.}\ \bibnamefont {Christodoulides}},\ }\href@noop {} {\bibfield
  {journal} {\bibinfo  {journal} {Phys. Rev. Lett.}\ }\textbf {\bibinfo
  {volume} {103}},\ \bibinfo {pages} {093902} (\bibinfo {year}
  {2009})}\BibitemShut {NoStop}%
\bibitem [{\citenamefont {R\"uter}\ \emph {et~al.}(2010)\citenamefont
  {R\"uter}, \citenamefont {Makris}, \citenamefont {El-Ganainy}, \citenamefont
  {Christodoulides}, \citenamefont {Segev},\ and\ \citenamefont
  {Kip}}]{RuterNatPhys2010}%
  \BibitemOpen
  \bibfield  {author} {\bibinfo {author} {\bibfnamefont {C.~E.}\ \bibnamefont
  {R\"uter}}, \bibinfo {author} {\bibfnamefont {K.~G.}\ \bibnamefont {Makris}},
  \bibinfo {author} {\bibfnamefont {R.}~\bibnamefont {El-Ganainy}}, \bibinfo
  {author} {\bibfnamefont {D.~N.}\ \bibnamefont {Christodoulides}}, \bibinfo
  {author} {\bibfnamefont {M.}~\bibnamefont {Segev}}, \ and\ \bibinfo {author}
  {\bibfnamefont {D.}~\bibnamefont {Kip}},\ }\href@noop {} {\bibfield
  {journal} {\bibinfo  {journal} {Nat. Phys.}\ }\textbf {\bibinfo {volume}
  {6}},\ \bibinfo {pages} {192} (\bibinfo {year} {2010})}\BibitemShut {NoStop}%
\bibitem [{\citenamefont {Regensburger}\ \emph {et~al.}(2012)\citenamefont
  {Regensburger}, \citenamefont {Bersch}, \citenamefont {Miri}, \citenamefont
  {Onishchukov}, \citenamefont {Christodoulides},\ and\ \citenamefont
  {Peschel}}]{RegensburgerNature2010}%
  \BibitemOpen
  \bibfield  {author} {\bibinfo {author} {\bibfnamefont {A.}~\bibnamefont
  {Regensburger}}, \bibinfo {author} {\bibfnamefont {C.}~\bibnamefont
  {Bersch}}, \bibinfo {author} {\bibfnamefont {M.-A.}\ \bibnamefont {Miri}},
  \bibinfo {author} {\bibfnamefont {G.}~\bibnamefont {Onishchukov}}, \bibinfo
  {author} {\bibfnamefont {D.~N.}\ \bibnamefont {Christodoulides}}, \ and\
  \bibinfo {author} {\bibfnamefont {U.}~\bibnamefont {Peschel}},\ }\href@noop
  {} {\bibfield  {journal} {\bibinfo  {journal} {Nature (London)}\ }\textbf
  {\bibinfo {volume} {488}},\ \bibinfo {pages} {167} (\bibinfo {year}
  {2012})}\BibitemShut {NoStop}%
\bibitem [{\citenamefont {Peng}\ \emph {et~al.}(2014)\citenamefont {Peng},
  \citenamefont {{\"O}zdemir}, \citenamefont {Lei}, \citenamefont {Monifi},
  \citenamefont {Gianfreda}, \citenamefont {Long}, \citenamefont {Fan},
  \citenamefont {Nori}, \citenamefont {Bender},\ and\ \citenamefont
  {Yang}}]{PengNatPhys2014}%
  \BibitemOpen
  \bibfield  {author} {\bibinfo {author} {\bibfnamefont {B.}~\bibnamefont
  {Peng}}, \bibinfo {author} {\bibfnamefont {{\c S}.~K.}\ \bibnamefont
  {{\"O}zdemir}}, \bibinfo {author} {\bibfnamefont {F.}~\bibnamefont {Lei}},
  \bibinfo {author} {\bibfnamefont {F.}~\bibnamefont {Monifi}}, \bibinfo
  {author} {\bibfnamefont {M.}~\bibnamefont {Gianfreda}}, \bibinfo {author}
  {\bibfnamefont {G.~L.}\ \bibnamefont {Long}}, \bibinfo {author}
  {\bibfnamefont {S.}~\bibnamefont {Fan}}, \bibinfo {author} {\bibfnamefont
  {F.}~\bibnamefont {Nori}}, \bibinfo {author} {\bibfnamefont {C.~M.}\
  \bibnamefont {Bender}}, \ and\ \bibinfo {author} {\bibfnamefont
  {L.}~\bibnamefont {Yang}},\ }\href@noop {} {\bibfield  {journal} {\bibinfo
  {journal} {Nature (London)}\ }\textbf {\bibinfo {volume} {10}},\ \bibinfo
  {pages} {394} (\bibinfo {year} {2014})}\BibitemShut {NoStop}%
\bibitem [{\citenamefont {Feng}\ \emph {et~al.}(2014)\citenamefont {Feng},
  \citenamefont {Wong}, \citenamefont {Ma}, \citenamefont {Wang},\ and\
  \citenamefont {Zhang}}]{FengScience2014}%
  \BibitemOpen
  \bibfield  {author} {\bibinfo {author} {\bibfnamefont {L.}~\bibnamefont
  {Feng}}, \bibinfo {author} {\bibfnamefont {Z.~J.}\ \bibnamefont {Wong}},
  \bibinfo {author} {\bibfnamefont {R.-M.}\ \bibnamefont {Ma}}, \bibinfo
  {author} {\bibfnamefont {Y.}~\bibnamefont {Wang}}, \ and\ \bibinfo {author}
  {\bibfnamefont {X.}~\bibnamefont {Zhang}},\ }\href@noop {} {\bibfield
  {journal} {\bibinfo  {journal} {Science}\ }\textbf {\bibinfo {volume}
  {346}},\ \bibinfo {pages} {972} (\bibinfo {year} {2014})}\BibitemShut
  {NoStop}%
\bibitem [{\citenamefont {Hodaei}\ \emph {et~al.}(2014)\citenamefont {Hodaei},
  \citenamefont {Miri}, \citenamefont {Heinrich}, \citenamefont
  {Christodoulides},\ and\ \citenamefont {Khajavikhan}}]{HodaelScience2014}%
  \BibitemOpen
  \bibfield  {author} {\bibinfo {author} {\bibfnamefont {H.}~\bibnamefont
  {Hodaei}}, \bibinfo {author} {\bibfnamefont {M.-A.}\ \bibnamefont {Miri}},
  \bibinfo {author} {\bibfnamefont {M.}~\bibnamefont {Heinrich}}, \bibinfo
  {author} {\bibfnamefont {D.~N.}\ \bibnamefont {Christodoulides}}, \ and\
  \bibinfo {author} {\bibfnamefont {M.}~\bibnamefont {Khajavikhan}},\
  }\href@noop {} {\bibfield  {journal} {\bibinfo  {journal} {Science}\ }\textbf
  {\bibinfo {volume} {346}},\ \bibinfo {pages} {975} (\bibinfo {year}
  {2014})}\BibitemShut {NoStop}%
\bibitem [{\citenamefont {Zhen}\ \emph {et~al.}(2015)\citenamefont {Zhen},
  \citenamefont {Hsu}, \citenamefont {Igarashi}, \citenamefont {Lu},
  \citenamefont {Kaminer}, \citenamefont {Pick}, \citenamefont {Chua},
  \citenamefont {Joannopoulos},\ and\ \citenamefont {Solj{\u
  c}i{\'c}}}]{ZhenNature2015}%
  \BibitemOpen
  \bibfield  {author} {\bibinfo {author} {\bibfnamefont {B.}~\bibnamefont
  {Zhen}}, \bibinfo {author} {\bibfnamefont {C.~W.}\ \bibnamefont {Hsu}},
  \bibinfo {author} {\bibfnamefont {Y.}~\bibnamefont {Igarashi}}, \bibinfo
  {author} {\bibfnamefont {L.}~\bibnamefont {Lu}}, \bibinfo {author}
  {\bibfnamefont {I.}~\bibnamefont {Kaminer}}, \bibinfo {author} {\bibfnamefont
  {A.}~\bibnamefont {Pick}}, \bibinfo {author} {\bibfnamefont {S.-L.}\
  \bibnamefont {Chua}}, \bibinfo {author} {\bibfnamefont {J.~D.}\ \bibnamefont
  {Joannopoulos}}, \ and\ \bibinfo {author} {\bibfnamefont {M.}~\bibnamefont
  {Solj{\u c}i{\'c}}},\ }\href@noop {} {\bibfield  {journal} {\bibinfo
  {journal} {Nature (London)}\ }\textbf {\bibinfo {volume} {525}},\ \bibinfo
  {pages} {354} (\bibinfo {year} {2015})}\BibitemShut {NoStop}%
\bibitem [{\citenamefont {Zeuner}\ \emph {et~al.}(2015)\citenamefont {Zeuner},
  \citenamefont {Rechtsman}, \citenamefont {Plotnik}, \citenamefont {Lumer},
  \citenamefont {Nolte}, \citenamefont {Rudner}, \citenamefont {Segev},\ and\
  \citenamefont {Szameit}}]{ZeunerPRL2015}%
  \BibitemOpen
  \bibfield  {author} {\bibinfo {author} {\bibfnamefont {J.~M.}\ \bibnamefont
  {Zeuner}}, \bibinfo {author} {\bibfnamefont {M.~C.}\ \bibnamefont
  {Rechtsman}}, \bibinfo {author} {\bibfnamefont {Y.}~\bibnamefont {Plotnik}},
  \bibinfo {author} {\bibfnamefont {Y.}~\bibnamefont {Lumer}}, \bibinfo
  {author} {\bibfnamefont {S.}~\bibnamefont {Nolte}}, \bibinfo {author}
  {\bibfnamefont {M.~S.}\ \bibnamefont {Rudner}}, \bibinfo {author}
  {\bibfnamefont {M.}~\bibnamefont {Segev}}, \ and\ \bibinfo {author}
  {\bibfnamefont {A.}~\bibnamefont {Szameit}},\ }\href@noop {} {\bibfield
  {journal} {\bibinfo  {journal} {Phys. Rev. Lett.}\ }\textbf {\bibinfo
  {volume} {115}},\ \bibinfo {pages} {040402} (\bibinfo {year}
  {2015})}\BibitemShut {NoStop}%
\bibitem [{\citenamefont {Poli}\ \emph {et~al.}(2015)\citenamefont {Poli},
  \citenamefont {Bellec}, \citenamefont {Kuhl}, \citenamefont {Mortessagne},\
  and\ \citenamefont {Schomerus}}]{PoliNatComm2015}%
  \BibitemOpen
  \bibfield  {author} {\bibinfo {author} {\bibfnamefont {C.}~\bibnamefont
  {Poli}}, \bibinfo {author} {\bibfnamefont {M.}~\bibnamefont {Bellec}},
  \bibinfo {author} {\bibfnamefont {U.}~\bibnamefont {Kuhl}}, \bibinfo {author}
  {\bibfnamefont {F.}~\bibnamefont {Mortessagne}}, \ and\ \bibinfo {author}
  {\bibfnamefont {H.}~\bibnamefont {Schomerus}},\ }\href@noop {} {\bibfield
  {journal} {\bibinfo  {journal} {Nat. Comm.}\ }\textbf {\bibinfo {volume}
  {6}},\ \bibinfo {pages} {6710} (\bibinfo {year} {2015})}\BibitemShut
  {NoStop}%
\bibitem [{\citenamefont {Doppler}\ \emph {et~al.}(2016)\citenamefont
  {Doppler}, \citenamefont {Mailybaev}, \citenamefont {B{\"o}hm}, \citenamefont
  {Kuhl}, \citenamefont {Girschik}, \citenamefont {Libisch}, \citenamefont
  {Milburn}, \citenamefont {Rabl}, \citenamefont {Moiseyev},\ and\
  \citenamefont {Rotter}}]{DopplerNature2016}%
  \BibitemOpen
  \bibfield  {author} {\bibinfo {author} {\bibfnamefont {J.}~\bibnamefont
  {Doppler}}, \bibinfo {author} {\bibfnamefont {A.~A.}\ \bibnamefont
  {Mailybaev}}, \bibinfo {author} {\bibfnamefont {J.}~\bibnamefont {B{\"o}hm}},
  \bibinfo {author} {\bibfnamefont {U.}~\bibnamefont {Kuhl}}, \bibinfo {author}
  {\bibfnamefont {A.}~\bibnamefont {Girschik}}, \bibinfo {author}
  {\bibfnamefont {F.}~\bibnamefont {Libisch}}, \bibinfo {author} {\bibfnamefont
  {T.~J.}\ \bibnamefont {Milburn}}, \bibinfo {author} {\bibfnamefont
  {P.}~\bibnamefont {Rabl}}, \bibinfo {author} {\bibfnamefont {N.}~\bibnamefont
  {Moiseyev}}, \ and\ \bibinfo {author} {\bibfnamefont {S.}~\bibnamefont
  {Rotter}},\ }\href@noop {} {\bibfield  {journal} {\bibinfo  {journal} {Nature
  (London)}\ }\textbf {\bibinfo {volume} {537}},\ \bibinfo {pages} {76}
  (\bibinfo {year} {2016})}\BibitemShut {NoStop}%
\bibitem [{\citenamefont {Bender}\ \emph {et~al.}(2013)\citenamefont {Bender},
  \citenamefont {Berntson}, \citenamefont {Parker},\ and\ \citenamefont
  {Samuel}}]{BenderAJP2013}%
  \BibitemOpen
  \bibfield  {author} {\bibinfo {author} {\bibfnamefont {C.~M.}\ \bibnamefont
  {Bender}}, \bibinfo {author} {\bibfnamefont {B.~K.}\ \bibnamefont
  {Berntson}}, \bibinfo {author} {\bibfnamefont {D.}~\bibnamefont {Parker}}, \
  and\ \bibinfo {author} {\bibfnamefont {E.}~\bibnamefont {Samuel}},\
  }\href@noop {} {\bibfield  {journal} {\bibinfo  {journal} {Am. J. Phys.}\
  }\textbf {\bibinfo {volume} {81}},\ \bibinfo {pages} {173} (\bibinfo {year}
  {2013})}\BibitemShut {NoStop}%
\bibitem [{\citenamefont {Li}\ \emph {et~al.}(2019)\citenamefont {Li},
  \citenamefont {Harter}, \citenamefont {Liu}, \citenamefont {de~Melo},
  \citenamefont {Joglekar},\ and\ \citenamefont {Luo}}]{LiNatComm2019}%
  \BibitemOpen
  \bibfield  {author} {\bibinfo {author} {\bibfnamefont {J.}~\bibnamefont
  {Li}}, \bibinfo {author} {\bibfnamefont {A.~K.}\ \bibnamefont {Harter}},
  \bibinfo {author} {\bibfnamefont {J.}~\bibnamefont {Liu}}, \bibinfo {author}
  {\bibfnamefont {L.}~\bibnamefont {de~Melo}}, \bibinfo {author} {\bibfnamefont
  {Y.~N.}\ \bibnamefont {Joglekar}}, \ and\ \bibinfo {author} {\bibfnamefont
  {L.}~\bibnamefont {Luo}},\ }\href@noop {} {\bibfield  {journal} {\bibinfo
  {journal} {Nat. Comm.}\ }\textbf {\bibinfo {volume} {10}},\ \bibinfo {pages}
  {855} (\bibinfo {year} {2019})}\BibitemShut {NoStop}%
\bibitem [{\citenamefont {Dal~Negro}\ \emph {et~al.}(2003)\citenamefont
  {Dal~Negro}, \citenamefont {Oton}, \citenamefont {Gaburro}, \citenamefont
  {Pavesi}, \citenamefont {Johnson}, \citenamefont {Lagendijk}, \citenamefont
  {Righini}, \citenamefont {Colocci},\ and\ \citenamefont
  {Wiersma}}]{NegroPRL2003}%
  \BibitemOpen
  \bibfield  {author} {\bibinfo {author} {\bibfnamefont {L.}~\bibnamefont
  {Dal~Negro}}, \bibinfo {author} {\bibfnamefont {C.~J.}\ \bibnamefont {Oton}},
  \bibinfo {author} {\bibfnamefont {Z.}~\bibnamefont {Gaburro}}, \bibinfo
  {author} {\bibfnamefont {L.}~\bibnamefont {Pavesi}}, \bibinfo {author}
  {\bibfnamefont {P.}~\bibnamefont {Johnson}}, \bibinfo {author} {\bibfnamefont
  {A.}~\bibnamefont {Lagendijk}}, \bibinfo {author} {\bibfnamefont
  {R.}~\bibnamefont {Righini}}, \bibinfo {author} {\bibfnamefont
  {M.}~\bibnamefont {Colocci}}, \ and\ \bibinfo {author} {\bibfnamefont
  {D.~S.}\ \bibnamefont {Wiersma}},\ }\href@noop {} {\bibfield  {journal}
  {\bibinfo  {journal} {Phys. Rev. Lett.}\ }\textbf {\bibinfo {volume} {90}},\
  \bibinfo {pages} {055501} (\bibinfo {year} {2003})}\BibitemShut {NoStop}%
\bibitem [{\citenamefont {Lahini}\ \emph {et~al.}(2009)\citenamefont {Lahini},
  \citenamefont {Pugatch}, \citenamefont {Pozzi}, \citenamefont {Sorel},
  \citenamefont {Morandotti}, \citenamefont {Davidson},\ and\ \citenamefont
  {Silberberg}}]{LahiniPRL2009}%
  \BibitemOpen
  \bibfield  {author} {\bibinfo {author} {\bibfnamefont {Y.}~\bibnamefont
  {Lahini}}, \bibinfo {author} {\bibfnamefont {R.}~\bibnamefont {Pugatch}},
  \bibinfo {author} {\bibfnamefont {F.}~\bibnamefont {Pozzi}}, \bibinfo
  {author} {\bibfnamefont {M.}~\bibnamefont {Sorel}}, \bibinfo {author}
  {\bibfnamefont {R.}~\bibnamefont {Morandotti}}, \bibinfo {author}
  {\bibfnamefont {N.}~\bibnamefont {Davidson}}, \ and\ \bibinfo {author}
  {\bibfnamefont {Y.}~\bibnamefont {Silberberg}},\ }\href@noop {} {\bibfield
  {journal} {\bibinfo  {journal} {Phys. Rev. Lett.}\ }\textbf {\bibinfo
  {volume} {103}},\ \bibinfo {pages} {013901} (\bibinfo {year}
  {2009})}\BibitemShut {NoStop}%
\bibitem [{\citenamefont {Kraus}\ \emph {et~al.}(2012)\citenamefont {Kraus},
  \citenamefont {Lahini}, \citenamefont {Ringel}, \citenamefont {Verbin},\ and\
  \citenamefont {Zilberberg}}]{KrausPRL2012}%
  \BibitemOpen
  \bibfield  {author} {\bibinfo {author} {\bibfnamefont {Y.~E.}\ \bibnamefont
  {Kraus}}, \bibinfo {author} {\bibfnamefont {Y.}~\bibnamefont {Lahini}},
  \bibinfo {author} {\bibfnamefont {Z.}~\bibnamefont {Ringel}}, \bibinfo
  {author} {\bibfnamefont {M.}~\bibnamefont {Verbin}}, \ and\ \bibinfo {author}
  {\bibfnamefont {O.}~\bibnamefont {Zilberberg}},\ }\href@noop {} {\bibfield
  {journal} {\bibinfo  {journal} {Phys. Rev. Lett.}\ }\textbf {\bibinfo
  {volume} {109}},\ \bibinfo {pages} {106402} (\bibinfo {year}
  {2012})}\BibitemShut {NoStop}%
\bibitem [{\citenamefont {Verbin}\ \emph {et~al.}(2013)\citenamefont {Verbin},
  \citenamefont {Zilberberg}, \citenamefont {Kraus}, \citenamefont {Lahini},\
  and\ \citenamefont {Silberberg}}]{VerbinPRL2013}%
  \BibitemOpen
  \bibfield  {author} {\bibinfo {author} {\bibfnamefont {M.}~\bibnamefont
  {Verbin}}, \bibinfo {author} {\bibfnamefont {O.}~\bibnamefont {Zilberberg}},
  \bibinfo {author} {\bibfnamefont {Y.~E.}\ \bibnamefont {Kraus}}, \bibinfo
  {author} {\bibfnamefont {Y.}~\bibnamefont {Lahini}}, \ and\ \bibinfo {author}
  {\bibfnamefont {Y.}~\bibnamefont {Silberberg}},\ }\href@noop {} {\bibfield
  {journal} {\bibinfo  {journal} {Phys. Rev. Lett.}\ }\textbf {\bibinfo
  {volume} {110}},\ \bibinfo {pages} {076403} (\bibinfo {year}
  {2013})}\BibitemShut {NoStop}%
\bibitem [{\citenamefont {Verbin}\ \emph {et~al.}(2015)\citenamefont {Verbin},
  \citenamefont {Zilberberg}, \citenamefont {Lahini}, \citenamefont {Kraus},\
  and\ \citenamefont {Silberberg}}]{VerbinPRB2015}%
  \BibitemOpen
  \bibfield  {author} {\bibinfo {author} {\bibfnamefont {M.}~\bibnamefont
  {Verbin}}, \bibinfo {author} {\bibfnamefont {O.}~\bibnamefont {Zilberberg}},
  \bibinfo {author} {\bibfnamefont {Y.}~\bibnamefont {Lahini}}, \bibinfo
  {author} {\bibfnamefont {Y.~E.}\ \bibnamefont {Kraus}}, \ and\ \bibinfo
  {author} {\bibfnamefont {Y.}~\bibnamefont {Silberberg}},\ }\href@noop {}
  {\bibfield  {journal} {\bibinfo  {journal} {Phys. Rev. B}\ }\textbf {\bibinfo
  {volume} {91}},\ \bibinfo {pages} {064201} (\bibinfo {year}
  {2015})}\BibitemShut {NoStop}%
\bibitem [{\citenamefont {Roati}\ \emph {et~al.}(2008)\citenamefont {Roati},
  \citenamefont {D'Errico}, \citenamefont {Fallani}, \citenamefont {Fattori},
  \citenamefont {Fort}, \citenamefont {Zaccanti}, \citenamefont {Modugno},
  \citenamefont {Modugno},\ and\ \citenamefont {Inguscio}}]{RoatiNature2008}%
  \BibitemOpen
  \bibfield  {author} {\bibinfo {author} {\bibfnamefont {G.}~\bibnamefont
  {Roati}}, \bibinfo {author} {\bibfnamefont {C.}~\bibnamefont {D'Errico}},
  \bibinfo {author} {\bibfnamefont {L.}~\bibnamefont {Fallani}}, \bibinfo
  {author} {\bibfnamefont {M.}~\bibnamefont {Fattori}}, \bibinfo {author}
  {\bibfnamefont {C.}~\bibnamefont {Fort}}, \bibinfo {author} {\bibfnamefont
  {M.}~\bibnamefont {Zaccanti}}, \bibinfo {author} {\bibfnamefont
  {G.}~\bibnamefont {Modugno}}, \bibinfo {author} {\bibfnamefont
  {M.}~\bibnamefont {Modugno}}, \ and\ \bibinfo {author} {\bibfnamefont
  {M.}~\bibnamefont {Inguscio}},\ }\href@noop {} {\bibfield  {journal}
  {\bibinfo  {journal} {Nature (London)}\ }\textbf {\bibinfo {volume} {453}},\
  \bibinfo {pages} {895} (\bibinfo {year} {2008})}\BibitemShut {NoStop}%
\bibitem [{\citenamefont {Modugno}(2010)}]{ModugnoRPP2010}%
  \BibitemOpen
  \bibfield  {author} {\bibinfo {author} {\bibfnamefont {G.}~\bibnamefont
  {Modugno}},\ }\href@noop {} {\bibfield  {journal} {\bibinfo  {journal} {Rep.
  Prog. Phys.}\ }\textbf {\bibinfo {volume} {73}},\ \bibinfo {pages} {102401}
  (\bibinfo {year} {2010})}\BibitemShut {NoStop}%
\bibitem [{\citenamefont {Anderson}(1958)}]{AndersonPR1958}%
  \BibitemOpen
  \bibfield  {author} {\bibinfo {author} {\bibfnamefont {P.~W.}\ \bibnamefont
  {Anderson}},\ }\href@noop {} {\bibfield  {journal} {\bibinfo  {journal}
  {Phys. Rev.}\ }\textbf {\bibinfo {volume} {109}},\ \bibinfo {pages} {1492}
  (\bibinfo {year} {1958})}\BibitemShut {NoStop}%
\bibitem [{\citenamefont {Aubry}\ and\ \citenamefont
  {Andr{\'e}}(1980)}]{AubryAIPS1980}%
  \BibitemOpen
  \bibfield  {author} {\bibinfo {author} {\bibfnamefont {S.}~\bibnamefont
  {Aubry}}\ and\ \bibinfo {author} {\bibfnamefont {G.}~\bibnamefont
  {Andr{\'e}}},\ }\href@noop {} {\bibfield  {journal} {\bibinfo  {journal}
  {Ann. Isr. Phys. Soc.}\ }\textbf {\bibinfo {volume} {3}},\ \bibinfo {pages}
  {18} (\bibinfo {year} {1980})}\BibitemShut {NoStop}%
\bibitem [{\citenamefont {Biddle}\ and\ \citenamefont
  {Das~Sarma}(2010)}]{BiddlePRL2010}%
  \BibitemOpen
  \bibfield  {author} {\bibinfo {author} {\bibfnamefont {J.}~\bibnamefont
  {Biddle}}\ and\ \bibinfo {author} {\bibfnamefont {S.}~\bibnamefont
  {Das~Sarma}},\ }\href@noop {} {\bibfield  {journal} {\bibinfo  {journal}
  {Phys. Rev. Lett.}\ }\textbf {\bibinfo {volume} {104}},\ \bibinfo {pages}
  {070601} (\bibinfo {year} {2010})}\BibitemShut {NoStop}%
\bibitem [{\citenamefont {Cai}\ \emph {et~al.}(2013)\citenamefont {Cai},
  \citenamefont {Lang}, \citenamefont {Chen},\ and\ \citenamefont
  {Wang}}]{CaiPRL2013}%
  \BibitemOpen
  \bibfield  {author} {\bibinfo {author} {\bibfnamefont {X.}~\bibnamefont
  {Cai}}, \bibinfo {author} {\bibfnamefont {L.-J.}\ \bibnamefont {Lang}},
  \bibinfo {author} {\bibfnamefont {S.}~\bibnamefont {Chen}}, \ and\ \bibinfo
  {author} {\bibfnamefont {Y.}~\bibnamefont {Wang}},\ }\href@noop {} {\bibfield
   {journal} {\bibinfo  {journal} {Phys. Rev. Lett.}\ }\textbf {\bibinfo
  {volume} {110}},\ \bibinfo {pages} {176403} (\bibinfo {year}
  {2013})}\BibitemShut {NoStop}%
\bibitem [{\citenamefont {DeGottardi}\ \emph {et~al.}(2013)\citenamefont
  {DeGottardi}, \citenamefont {Sen},\ and\ \citenamefont
  {Vishveshwara}}]{DeGottardiPRL2013}%
  \BibitemOpen
  \bibfield  {author} {\bibinfo {author} {\bibfnamefont {W.}~\bibnamefont
  {DeGottardi}}, \bibinfo {author} {\bibfnamefont {D.}~\bibnamefont {Sen}}, \
  and\ \bibinfo {author} {\bibfnamefont {S.}~\bibnamefont {Vishveshwara}},\
  }\href@noop {} {\bibfield  {journal} {\bibinfo  {journal} {Phys. Rev. Lett.}\
  }\textbf {\bibinfo {volume} {110}},\ \bibinfo {pages} {146404} (\bibinfo
  {year} {2013})}\BibitemShut {NoStop}%
\bibitem [{\citenamefont {Ganeshan}\ \emph {et~al.}(2015)\citenamefont
  {Ganeshan}, \citenamefont {Pixley},\ and\ \citenamefont
  {Das~Sarma}}]{GaneshanPRL2015}%
  \BibitemOpen
  \bibfield  {author} {\bibinfo {author} {\bibfnamefont {S.}~\bibnamefont
  {Ganeshan}}, \bibinfo {author} {\bibfnamefont {J.~H.}\ \bibnamefont
  {Pixley}}, \ and\ \bibinfo {author} {\bibfnamefont {S.}~\bibnamefont
  {Das~Sarma}},\ }\href@noop {} {\bibfield  {journal} {\bibinfo  {journal}
  {Phys. Rev. Lett.}\ }\textbf {\bibinfo {volume} {114}},\ \bibinfo {pages}
  {146601} (\bibinfo {year} {2015})}\BibitemShut {NoStop}%
\bibitem [{\citenamefont {Liu}\ \emph {et~al.}(2015)\citenamefont {Liu},
  \citenamefont {Ghosh},\ and\ \citenamefont {Chong}}]{LiuPRB2015}%
  \BibitemOpen
  \bibfield  {author} {\bibinfo {author} {\bibfnamefont {F.}~\bibnamefont
  {Liu}}, \bibinfo {author} {\bibfnamefont {S.}~\bibnamefont {Ghosh}}, \ and\
  \bibinfo {author} {\bibfnamefont {Y.~D.}\ \bibnamefont {Chong}},\ }\href@noop
  {} {\bibfield  {journal} {\bibinfo  {journal} {Phys. Rev. B}\ }\textbf
  {\bibinfo {volume} {91}},\ \bibinfo {pages} {014108} (\bibinfo {year}
  {2015})}\BibitemShut {NoStop}%
\bibitem [{\citenamefont {Wang}\ \emph {et~al.}(2016)\citenamefont {Wang},
  \citenamefont {Liu}, \citenamefont {Xianlong},\ and\ \citenamefont
  {Hu}}]{WangPRB2016}%
  \BibitemOpen
  \bibfield  {author} {\bibinfo {author} {\bibfnamefont {J.}~\bibnamefont
  {Wang}}, \bibinfo {author} {\bibfnamefont {X.-J.}\ \bibnamefont {Liu}},
  \bibinfo {author} {\bibfnamefont {G.}~\bibnamefont {Xianlong}}, \ and\
  \bibinfo {author} {\bibfnamefont {H.}~\bibnamefont {Hu}},\ }\href@noop {}
  {\bibfield  {journal} {\bibinfo  {journal} {Phys. Rev. B}\ }\textbf {\bibinfo
  {volume} {93}},\ \bibinfo {pages} {104504} (\bibinfo {year}
  {2016})}\BibitemShut {NoStop}%
\bibitem [{\citenamefont {Cao}\ \emph {et~al.}(2016)\citenamefont {Cao},
  \citenamefont {Gao}, \citenamefont {Liu},\ and\ \citenamefont
  {Hu}}]{CaoPRA2016}%
  \BibitemOpen
  \bibfield  {author} {\bibinfo {author} {\bibfnamefont {Y.}~\bibnamefont
  {Cao}}, \bibinfo {author} {\bibfnamefont {X.}~\bibnamefont {Gao}}, \bibinfo
  {author} {\bibfnamefont {X.-J.}\ \bibnamefont {Liu}}, \ and\ \bibinfo
  {author} {\bibfnamefont {H.}~\bibnamefont {Hu}},\ }\href@noop {} {\bibfield
  {journal} {\bibinfo  {journal} {Phys. Rev. A}\ }\textbf {\bibinfo {volume}
  {93}},\ \bibinfo {pages} {043621} (\bibinfo {year} {2016})}\BibitemShut
  {NoStop}%
\bibitem [{\citenamefont {Cestari}\ \emph {et~al.}(2016)\citenamefont
  {Cestari}, \citenamefont {Foerster},\ and\ \citenamefont
  {Gusm\~ao}}]{CestariPRB2016}%
  \BibitemOpen
  \bibfield  {author} {\bibinfo {author} {\bibfnamefont {J.~C.~C.}\
  \bibnamefont {Cestari}}, \bibinfo {author} {\bibfnamefont {A.}~\bibnamefont
  {Foerster}}, \ and\ \bibinfo {author} {\bibfnamefont {M.~A.}\ \bibnamefont
  {Gusm\~ao}},\ }\href@noop {} {\bibfield  {journal} {\bibinfo  {journal}
  {Phys. Rev. B}\ }\textbf {\bibinfo {volume} {93}},\ \bibinfo {pages} {205441}
  (\bibinfo {year} {2016})}\BibitemShut {NoStop}%
\bibitem [{\citenamefont {Zeng}\ \emph {et~al.}(2016)\citenamefont {Zeng},
  \citenamefont {Chen},\ and\ \citenamefont {L\"u}}]{ZengPRB2016}%
  \BibitemOpen
  \bibfield  {author} {\bibinfo {author} {\bibfnamefont {Q.-B.}\ \bibnamefont
  {Zeng}}, \bibinfo {author} {\bibfnamefont {S.}~\bibnamefont {Chen}}, \ and\
  \bibinfo {author} {\bibfnamefont {R.}~\bibnamefont {L\"u}},\ }\href@noop {}
  {\bibfield  {journal} {\bibinfo  {journal} {Phys. Rev. B}\ }\textbf {\bibinfo
  {volume} {94}},\ \bibinfo {pages} {125408} (\bibinfo {year}
  {2016})}\BibitemShut {NoStop}%
\bibitem [{\citenamefont {Bai}\ \emph {et~al.}(2018)\citenamefont {Bai},
  \citenamefont {Wang}, \citenamefont {Liu}, \citenamefont {Xiong},
  \citenamefont {Deng},\ and\ \citenamefont {Hu}}]{BaiPRA2018}%
  \BibitemOpen
  \bibfield  {author} {\bibinfo {author} {\bibfnamefont {X.-D.}\ \bibnamefont
  {Bai}}, \bibinfo {author} {\bibfnamefont {J.}~\bibnamefont {Wang}}, \bibinfo
  {author} {\bibfnamefont {X.-J.}\ \bibnamefont {Liu}}, \bibinfo {author}
  {\bibfnamefont {J.}~\bibnamefont {Xiong}}, \bibinfo {author} {\bibfnamefont
  {F.-G.}\ \bibnamefont {Deng}}, \ and\ \bibinfo {author} {\bibfnamefont
  {H.}~\bibnamefont {Hu}},\ }\href@noop {} {\bibfield  {journal} {\bibinfo
  {journal} {Phys. Rev. A}\ }\textbf {\bibinfo {volume} {98}},\ \bibinfo
  {pages} {023627} (\bibinfo {year} {2018})}\BibitemShut {NoStop}%
\bibitem [{\citenamefont {Yao}\ \emph {et~al.}(2019)\citenamefont {Yao},
  \citenamefont {Khoudli}, \citenamefont {Bresque},\ and\ \citenamefont
  {Sanchez-Palencia}}]{Sanchez-PalenciaPRL2019}%
  \BibitemOpen
  \bibfield  {author} {\bibinfo {author} {\bibfnamefont {H.}~\bibnamefont
  {Yao}}, \bibinfo {author} {\bibfnamefont {H.}~\bibnamefont {Khoudli}},
  \bibinfo {author} {\bibfnamefont {L.}~\bibnamefont {Bresque}}, \ and\
  \bibinfo {author} {\bibfnamefont {L.}~\bibnamefont {Sanchez-Palencia}},\
  }\href@noop {} {\bibfield  {journal} {\bibinfo  {journal} {Phys. Rev. Lett.}\
  }\textbf {\bibinfo {volume} {123}},\ \bibinfo {pages} {070405} (\bibinfo
  {year} {2019})}\BibitemShut {NoStop}%
\bibitem [{\citenamefont {Yao}\ \emph {et~al.}(2020)\citenamefont {Yao},
  \citenamefont {Giamarchi},\ and\ \citenamefont
  {Sanchez-Palencia}}]{Sanchez-PalenciaPRL2020}%
  \BibitemOpen
  \bibfield  {author} {\bibinfo {author} {\bibfnamefont {H.}~\bibnamefont
  {Yao}}, \bibinfo {author} {\bibfnamefont {T.}~\bibnamefont {Giamarchi}}, \
  and\ \bibinfo {author} {\bibfnamefont {L.}~\bibnamefont {Sanchez-Palencia}},\
  }\href@noop {} {\bibfield  {journal} {\bibinfo  {journal} {Phys. Rev. Lett.}\
  }\textbf {\bibinfo {volume} {125}},\ \bibinfo {pages} {060401} (\bibinfo
  {year} {2020})}\BibitemShut {NoStop}%
\bibitem [{\citenamefont {An}\ \emph {et~al.}()\citenamefont {An},
  \citenamefont {Padavi{\'c}}, \citenamefont {Meier}, \citenamefont {Hegde},
  \citenamefont {Ganeshan}, \citenamefont {Pixley}, \citenamefont
  {Vishveshwara},\ and\ \citenamefont {Gadway}}]{GadwayPreprint2020}%
  \BibitemOpen
  \bibfield  {author} {\bibinfo {author} {\bibfnamefont {F.~A.}\ \bibnamefont
  {An}}, \bibinfo {author} {\bibfnamefont {K.}~\bibnamefont {Padavi{\'c}}},
  \bibinfo {author} {\bibfnamefont {E.~J.}\ \bibnamefont {Meier}}, \bibinfo
  {author} {\bibfnamefont {S.}~\bibnamefont {Hegde}}, \bibinfo {author}
  {\bibfnamefont {S.}~\bibnamefont {Ganeshan}}, \bibinfo {author}
  {\bibfnamefont {J.}~\bibnamefont {Pixley}}, \bibinfo {author} {\bibfnamefont
  {S.}~\bibnamefont {Vishveshwara}}, \ and\ \bibinfo {author} {\bibfnamefont
  {B.}~\bibnamefont {Gadway}},\ }\href@noop {} {}\bibinfo {note} {``Observation
  of tunable mobility edges in generalized Aubry-Andr{\'e} lattices",
  arXiv:2007.01393 (2020)}\BibitemShut {NoStop}%
\bibitem [{\citenamefont {Zeng}\ \emph {et~al.}(2017)\citenamefont {Zeng},
  \citenamefont {Chen},\ and\ \citenamefont {L\"u}}]{ZengPRA2017}%
  \BibitemOpen
  \bibfield  {author} {\bibinfo {author} {\bibfnamefont {Q.-B.}\ \bibnamefont
  {Zeng}}, \bibinfo {author} {\bibfnamefont {S.}~\bibnamefont {Chen}}, \ and\
  \bibinfo {author} {\bibfnamefont {R.}~\bibnamefont {L\"u}},\ }\href@noop {}
  {\bibfield  {journal} {\bibinfo  {journal} {Phys. Rev. A}\ }\textbf {\bibinfo
  {volume} {95}},\ \bibinfo {pages} {062118} (\bibinfo {year}
  {2017})}\BibitemShut {NoStop}%
\bibitem [{\citenamefont {Liu}\ \emph {et~al.}(2020)\citenamefont {Liu},
  \citenamefont {Guo}, \citenamefont {Pu},\ and\ \citenamefont
  {Longhi}}]{LiuPRB2020}%
  \BibitemOpen
  \bibfield  {author} {\bibinfo {author} {\bibfnamefont {T.}~\bibnamefont
  {Liu}}, \bibinfo {author} {\bibfnamefont {H.}~\bibnamefont {Guo}}, \bibinfo
  {author} {\bibfnamefont {Y.}~\bibnamefont {Pu}}, \ and\ \bibinfo {author}
  {\bibfnamefont {S.}~\bibnamefont {Longhi}},\ }\href@noop {} {\bibfield
  {journal} {\bibinfo  {journal} {Phys. Rev. B}\ }\textbf {\bibinfo {volume}
  {102}},\ \bibinfo {pages} {024205} (\bibinfo {year} {2020})}\BibitemShut
  {NoStop}%
\bibitem [{\citenamefont {Zeng}\ \emph {et~al.}(2020)\citenamefont {Zeng},
  \citenamefont {Yang},\ and\ \citenamefont {Xu}}]{zengPRB2020}%
  \BibitemOpen
  \bibfield  {author} {\bibinfo {author} {\bibfnamefont {Q.-B.}\ \bibnamefont
  {Zeng}}, \bibinfo {author} {\bibfnamefont {Y.-B.}\ \bibnamefont {Yang}}, \
  and\ \bibinfo {author} {\bibfnamefont {Y.}~\bibnamefont {Xu}},\ }\href@noop
  {} {\bibfield  {journal} {\bibinfo  {journal} {Phys. Rev. B}\ }\textbf
  {\bibinfo {volume} {101}},\ \bibinfo {pages} {020201} (\bibinfo {year}
  {2020})}\BibitemShut {NoStop}%
\bibitem [{\citenamefont {Hatano}\ and\ \citenamefont
  {Nelson}(1996)}]{HatanoPRL1996}%
  \BibitemOpen
  \bibfield  {author} {\bibinfo {author} {\bibfnamefont {N.}~\bibnamefont
  {Hatano}}\ and\ \bibinfo {author} {\bibfnamefont {D.~R.}\ \bibnamefont
  {Nelson}},\ }\href@noop {} {\bibfield  {journal} {\bibinfo  {journal} {Phys.
  Rev. Lett.}\ }\textbf {\bibinfo {volume} {77}},\ \bibinfo {pages} {570}
  (\bibinfo {year} {1996})}\BibitemShut {NoStop}%
\bibitem [{\citenamefont {Hatano}\ and\ \citenamefont
  {Nelson}(1997)}]{HatanoPRB1997}%
  \BibitemOpen
  \bibfield  {author} {\bibinfo {author} {\bibfnamefont {N.}~\bibnamefont
  {Hatano}}\ and\ \bibinfo {author} {\bibfnamefont {D.~R.}\ \bibnamefont
  {Nelson}},\ }\href@noop {} {\bibfield  {journal} {\bibinfo  {journal} {Phys.
  Rev. B}\ }\textbf {\bibinfo {volume} {56}},\ \bibinfo {pages} {8651}
  (\bibinfo {year} {1997})}\BibitemShut {NoStop}%
\bibitem [{\citenamefont {Hatano}\ and\ \citenamefont
  {Nelson}(1998)}]{HatanoPRB1998}%
  \BibitemOpen
  \bibfield  {author} {\bibinfo {author} {\bibfnamefont {N.}~\bibnamefont
  {Hatano}}\ and\ \bibinfo {author} {\bibfnamefont {D.~R.}\ \bibnamefont
  {Nelson}},\ }\href@noop {} {\bibfield  {journal} {\bibinfo  {journal} {Phys.
  Rev. B}\ }\textbf {\bibinfo {volume} {58}},\ \bibinfo {pages} {8384}
  (\bibinfo {year} {1998})}\BibitemShut {NoStop}%
\bibitem [{\citenamefont {Hamazaki}\ \emph {et~al.}(2019)\citenamefont
  {Hamazaki}, \citenamefont {Kawabata},\ and\ \citenamefont
  {Ueda}}]{HamazakiPRL2019}%
  \BibitemOpen
  \bibfield  {author} {\bibinfo {author} {\bibfnamefont {R.}~\bibnamefont
  {Hamazaki}}, \bibinfo {author} {\bibfnamefont {K.}~\bibnamefont {Kawabata}},
  \ and\ \bibinfo {author} {\bibfnamefont {M.}~\bibnamefont {Ueda}},\
  }\href@noop {} {\bibfield  {journal} {\bibinfo  {journal} {Phys. Rev. Lett.}\
  }\textbf {\bibinfo {volume} {123}},\ \bibinfo {pages} {090603} (\bibinfo
  {year} {2019})}\BibitemShut {NoStop}%
\bibitem [{\citenamefont {Gong}\ \emph {et~al.}(2018)\citenamefont {Gong},
  \citenamefont {Ashida}, \citenamefont {Kawabata}, \citenamefont {Takasan},
  \citenamefont {Higashikawa},\ and\ \citenamefont {Ueda}}]{GongPRX2018}%
  \BibitemOpen
  \bibfield  {author} {\bibinfo {author} {\bibfnamefont {Z.}~\bibnamefont
  {Gong}}, \bibinfo {author} {\bibfnamefont {Y.}~\bibnamefont {Ashida}},
  \bibinfo {author} {\bibfnamefont {K.}~\bibnamefont {Kawabata}}, \bibinfo
  {author} {\bibfnamefont {K.}~\bibnamefont {Takasan}}, \bibinfo {author}
  {\bibfnamefont {S.}~\bibnamefont {Higashikawa}}, \ and\ \bibinfo {author}
  {\bibfnamefont {M.}~\bibnamefont {Ueda}},\ }\href@noop {} {\bibfield
  {journal} {\bibinfo  {journal} {Phys. Rev. X}\ }\textbf {\bibinfo {volume}
  {8}},\ \bibinfo {pages} {031079} (\bibinfo {year} {2018})}\BibitemShut
  {NoStop}%
\bibitem [{\citenamefont {Mej\'{\i}a-Cort\'es}\ and\ \citenamefont
  {Molina}(2015)}]{CristianPRA2015}%
  \BibitemOpen
  \bibfield  {author} {\bibinfo {author} {\bibfnamefont {C.}~\bibnamefont
  {Mej\'{\i}a-Cort\'es}}\ and\ \bibinfo {author} {\bibfnamefont {M.~I.}\
  \bibnamefont {Molina}},\ }\href@noop {} {\bibfield  {journal} {\bibinfo
  {journal} {Phys. Rev. A}\ }\textbf {\bibinfo {volume} {91}},\ \bibinfo
  {pages} {033815} (\bibinfo {year} {2015})}\BibitemShut {NoStop}%
\bibitem [{\citenamefont {Jovi{\'c}}\ \emph {et~al.}(2012)\citenamefont
  {Jovi{\'c}}, \citenamefont {Denz},\ and\ \citenamefont
  {Beli{\'c}}}]{JovicOL2012}%
  \BibitemOpen
  \bibfield  {author} {\bibinfo {author} {\bibfnamefont {D.~M.}\ \bibnamefont
  {Jovi{\'c}}}, \bibinfo {author} {\bibfnamefont {C.}~\bibnamefont {Denz}}, \
  and\ \bibinfo {author} {\bibfnamefont {M.~R.}\ \bibnamefont {Beli{\'c}}},\
  }\href@noop {} {\bibfield  {journal} {\bibinfo  {journal} {Opt. Lett.}\
  }\textbf {\bibinfo {volume} {37}},\ \bibinfo {pages} {4455} (\bibinfo {year}
  {2012})}\BibitemShut {NoStop}%
\bibitem [{\citenamefont {Joglekar}\ \emph {et~al.}(2010)\citenamefont
  {Joglekar}, \citenamefont {Scott}, \citenamefont {Babbey},\ and\
  \citenamefont {Saxena}}]{JoglekarPRA2010}%
  \BibitemOpen
  \bibfield  {author} {\bibinfo {author} {\bibfnamefont {Y.~N.}\ \bibnamefont
  {Joglekar}}, \bibinfo {author} {\bibfnamefont {D.}~\bibnamefont {Scott}},
  \bibinfo {author} {\bibfnamefont {M.}~\bibnamefont {Babbey}}, \ and\ \bibinfo
  {author} {\bibfnamefont {A.}~\bibnamefont {Saxena}},\ }\href@noop {}
  {\bibfield  {journal} {\bibinfo  {journal} {Phys. Rev. A}\ }\textbf {\bibinfo
  {volume} {82}},\ \bibinfo {pages} {030103} (\bibinfo {year}
  {2010})}\BibitemShut {NoStop}%
\bibitem [{\citenamefont {Liang}\ \emph {et~al.}(2014)\citenamefont {Liang},
  \citenamefont {Scott},\ and\ \citenamefont {Joglekar}}]{LiangPRA2014}%
  \BibitemOpen
  \bibfield  {author} {\bibinfo {author} {\bibfnamefont {C.~H.}\ \bibnamefont
  {Liang}}, \bibinfo {author} {\bibfnamefont {D.~D.}\ \bibnamefont {Scott}}, \
  and\ \bibinfo {author} {\bibfnamefont {Y.~N.}\ \bibnamefont {Joglekar}},\
  }\href@noop {} {\bibfield  {journal} {\bibinfo  {journal} {Phys. Rev. A}\
  }\textbf {\bibinfo {volume} {89}},\ \bibinfo {pages} {030102} (\bibinfo
  {year} {2014})}\BibitemShut {NoStop}%
\bibitem [{\citenamefont {Yuce}(2015)}]{YucePLA2014}%
  \BibitemOpen
  \bibfield  {author} {\bibinfo {author} {\bibfnamefont {C.}~\bibnamefont
  {Yuce}},\ }\href@noop {} {\bibfield  {journal} {\bibinfo  {journal} {Phys.
  Lett. A}\ }\textbf {\bibinfo {volume} {378}},\ \bibinfo {pages} {2024}
  (\bibinfo {year} {2015})}\BibitemShut {NoStop}%
\bibitem [{\citenamefont {Joglekar}\ and\ \citenamefont
  {Saxena}(2011)}]{JoglekarPRA2011}%
  \BibitemOpen
  \bibfield  {author} {\bibinfo {author} {\bibfnamefont {Y.~N.}\ \bibnamefont
  {Joglekar}}\ and\ \bibinfo {author} {\bibfnamefont {A.}~\bibnamefont
  {Saxena}},\ }\href@noop {} {\bibfield  {journal} {\bibinfo  {journal} {Phys.
  Rev. A}\ }\textbf {\bibinfo {volume} {83}},\ \bibinfo {pages} {050101}
  (\bibinfo {year} {2011})}\BibitemShut {NoStop}%
\bibitem [{\citenamefont {Scott}\ and\ \citenamefont
  {Joglekar}(2011)}]{ScottPRA2011}%
  \BibitemOpen
  \bibfield  {author} {\bibinfo {author} {\bibfnamefont {D.~D.}\ \bibnamefont
  {Scott}}\ and\ \bibinfo {author} {\bibfnamefont {Y.~N.}\ \bibnamefont
  {Joglekar}},\ }\href@noop {} {\bibfield  {journal} {\bibinfo  {journal}
  {Phys. Rev. A}\ }\textbf {\bibinfo {volume} {83}},\ \bibinfo {pages} {050102}
  (\bibinfo {year} {2011})}\BibitemShut {NoStop}%
\bibitem [{\citenamefont {Harter}\ \emph {et~al.}(2016)\citenamefont {Harter},
  \citenamefont {Lee},\ and\ \citenamefont {Joglekar}}]{HarterPRA2016}%
  \BibitemOpen
  \bibfield  {author} {\bibinfo {author} {\bibfnamefont {A.~K.}\ \bibnamefont
  {Harter}}, \bibinfo {author} {\bibfnamefont {T.~E.}\ \bibnamefont {Lee}}, \
  and\ \bibinfo {author} {\bibfnamefont {Y.~N.}\ \bibnamefont {Joglekar}},\
  }\href@noop {} {\bibfield  {journal} {\bibinfo  {journal} {Phys. Rev. A}\
  }\textbf {\bibinfo {volume} {93}},\ \bibinfo {pages} {062101} (\bibinfo
  {year} {2016})}\BibitemShut {NoStop}%
\bibitem [{Sup()}]{Supp}%
  \BibitemOpen
  \href@noop {} {}\bibinfo {note} {See Supplemental Materials for the details
  of (I) analytical investigation of the system's symmetry, (II) $\langle \rm
  MIPR \rangle$ and $\langle r \rangle$ near the center of the spectrum, and
  (III) technical details of our multifractal analysis. (IV) a possible
  experimental realization of our model.}\BibitemShut {Stop}%
\bibitem [{\citenamefont {Sch{\"a}fer}\ and\ \citenamefont
  {Wegner}(1980)}]{Schaefer1980}%
  \BibitemOpen
  \bibfield  {author} {\bibinfo {author} {\bibfnamefont {L.}~\bibnamefont
  {Sch{\"a}fer}}\ and\ \bibinfo {author} {\bibfnamefont {F.}~\bibnamefont
  {Wegner}},\ }\href@noop {} {\bibfield  {journal} {\bibinfo  {journal}
  {Zeitschrift f{\"u}r Physik B Condensed Matter}\ }\textbf {\bibinfo {volume}
  {39}},\ \bibinfo {pages} {281} (\bibinfo {year} {1980})}\BibitemShut
  {NoStop}%
\bibitem [{\citenamefont {Zhang}\ and\ \citenamefont
  {Nelson}(2019)}]{Zhang2019}%
  \BibitemOpen
  \bibfield  {author} {\bibinfo {author} {\bibfnamefont {G.~H.}\ \bibnamefont
  {Zhang}}\ and\ \bibinfo {author} {\bibfnamefont {D.~R.}\ \bibnamefont
  {Nelson}},\ }\href@noop {} {\bibfield  {journal} {\bibinfo  {journal} {Phys.
  Rev. E}\ }\textbf {\bibinfo {volume} {100}},\ \bibinfo {pages} {052315}
  (\bibinfo {year} {2019})}\BibitemShut {NoStop}%
\bibitem [{\citenamefont {Oganesyan}\ and\ \citenamefont
  {Huse}(2007)}]{Oganesyan2007}%
  \BibitemOpen
  \bibfield  {author} {\bibinfo {author} {\bibfnamefont {V.}~\bibnamefont
  {Oganesyan}}\ and\ \bibinfo {author} {\bibfnamefont {D.~A.}\ \bibnamefont
  {Huse}},\ }\href@noop {} {\bibfield  {journal} {\bibinfo  {journal} {Phys.
  Rev. B}\ }\textbf {\bibinfo {volume} {75}},\ \bibinfo {pages} {155111}
  (\bibinfo {year} {2007})}\BibitemShut {NoStop}%
\bibitem [{\citenamefont {Pal}\ and\ \citenamefont {Huse}(2010)}]{Pal2010}%
  \BibitemOpen
  \bibfield  {author} {\bibinfo {author} {\bibfnamefont {A.}~\bibnamefont
  {Pal}}\ and\ \bibinfo {author} {\bibfnamefont {D.~A.}\ \bibnamefont {Huse}},\
  }\href@noop {} {\bibfield  {journal} {\bibinfo  {journal} {Phys. Rev. B}\
  }\textbf {\bibinfo {volume} {82}},\ \bibinfo {pages} {174411} (\bibinfo
  {year} {2010})}\BibitemShut {NoStop}%
\bibitem [{\citenamefont {Hiramoto}\ and\ \citenamefont
  {Kohmoto}(1989)}]{Hiramoto1989}%
  \BibitemOpen
  \bibfield  {author} {\bibinfo {author} {\bibfnamefont {H.}~\bibnamefont
  {Hiramoto}}\ and\ \bibinfo {author} {\bibfnamefont {M.}~\bibnamefont
  {Kohmoto}},\ }\href@noop {} {\bibfield  {journal} {\bibinfo  {journal}
  {Physical Review B}\ }\textbf {\bibinfo {volume} {40}},\ \bibinfo {pages}
  {8225} (\bibinfo {year} {1989})}\BibitemShut {NoStop}%
\bibitem [{\citenamefont {Takasu}\ \emph {et~al.}()\citenamefont {Takasu},
  \citenamefont {Yagami}, \citenamefont {Ashida}, \citenamefont {Hamazaki},
  \citenamefont {Kuno},\ and\ \citenamefont
  {Takahashi}}]{TakahashiPreprint2020}%
  \BibitemOpen
  \bibfield  {author} {\bibinfo {author} {\bibfnamefont {Y.}~\bibnamefont
  {Takasu}}, \bibinfo {author} {\bibfnamefont {T.}~\bibnamefont {Yagami}},
  \bibinfo {author} {\bibfnamefont {Y.}~\bibnamefont {Ashida}}, \bibinfo
  {author} {\bibfnamefont {R.}~\bibnamefont {Hamazaki}}, \bibinfo {author}
  {\bibfnamefont {Y.}~\bibnamefont {Kuno}}, \ and\ \bibinfo {author}
  {\bibfnamefont {Y.}~\bibnamefont {Takahashi}},\ }\href@noop {} {}\bibinfo
  {note} {``PT-symmetric non-Hermitian quantum many-body system using ultracold
  atoms in an optical lattice with controlled dissipation", arXiv:2004.05734
  (2020)}\BibitemShut {NoStop}%
\bibitem [{\citenamefont {Rosen}(2005)}]{Rosen2005}%
  \BibitemOpen
  \bibfield  {author} {\bibinfo {author} {\bibfnamefont {K.~H.}\ \bibnamefont
  {Rosen}},\ }\href@noop {} {\emph {\bibinfo {title} {Elementary Number Theory
  and Its Applications}}}\ (\bibinfo  {publisher} {Pearson/Addison Wesley,},\
  \bibinfo {year} {2005})\BibitemShut {NoStop}%
\bibitem [{\citenamefont {Aidelsburger}\ \emph {et~al.}(2011)\citenamefont
  {Aidelsburger}, \citenamefont {Atala}, \citenamefont {Nascimb\`ene},
  \citenamefont {Trotzky}, \citenamefont {Chen},\ and\ \citenamefont
  {Bloch}}]{BlochPRL2011}%
  \BibitemOpen
  \bibfield  {author} {\bibinfo {author} {\bibfnamefont {M.}~\bibnamefont
  {Aidelsburger}}, \bibinfo {author} {\bibfnamefont {M.}~\bibnamefont {Atala}},
  \bibinfo {author} {\bibfnamefont {S.}~\bibnamefont {Nascimb\`ene}}, \bibinfo
  {author} {\bibfnamefont {S.}~\bibnamefont {Trotzky}}, \bibinfo {author}
  {\bibfnamefont {Y.-A.}\ \bibnamefont {Chen}}, \ and\ \bibinfo {author}
  {\bibfnamefont {I.}~\bibnamefont {Bloch}},\ }\href@noop {} {\bibfield
  {journal} {\bibinfo  {journal} {Phys. Rev. Lett.}\ }\textbf {\bibinfo
  {volume} {107}},\ \bibinfo {pages} {255301} (\bibinfo {year}
  {2011})}\BibitemShut {NoStop}%
\bibitem [{\citenamefont {Jim\'enez-Garc\'{\i}a}\ \emph
  {et~al.}(2012)\citenamefont {Jim\'enez-Garc\'{\i}a}, \citenamefont {LeBlanc},
  \citenamefont {Williams}, \citenamefont {Beeler}, \citenamefont {Perry},\
  and\ \citenamefont {Spielman}}]{SpielmanPRL2012}%
  \BibitemOpen
  \bibfield  {author} {\bibinfo {author} {\bibfnamefont {K.}~\bibnamefont
  {Jim\'enez-Garc\'{\i}a}}, \bibinfo {author} {\bibfnamefont {L.~J.}\
  \bibnamefont {LeBlanc}}, \bibinfo {author} {\bibfnamefont {R.~A.}\
  \bibnamefont {Williams}}, \bibinfo {author} {\bibfnamefont {M.~C.}\
  \bibnamefont {Beeler}}, \bibinfo {author} {\bibfnamefont {A.~R.}\
  \bibnamefont {Perry}}, \ and\ \bibinfo {author} {\bibfnamefont {I.~B.}\
  \bibnamefont {Spielman}},\ }\href@noop {} {\bibfield  {journal} {\bibinfo
  {journal} {Phys. Rev. Lett.}\ }\textbf {\bibinfo {volume} {108}},\ \bibinfo
  {pages} {225303} (\bibinfo {year} {2012})}\BibitemShut {NoStop}%
\end{thebibliography}%

\end{document}